# AI-predicted PT-symmetric magnets

Hao Wu✉ and Daniel F. Agterberg

Department of Physics, University of Wisconsin–Milwaukee, Milwaukee, Wisconsin 53201, USA

**Abstract**

Parity-time-reversal-symmetric odd-parity antiferromagnetic (AFM1) materials are of interest for their symmetry-enabled quantum transport and optical effects. These materials host odd-parity terms in their band dispersion, leading to asymmetric energy bands and enabling responses such as the magnetopiezoelectric effect, nonreciprocal conductivity, and photocurrent generation. In addition, they may support a nonlinear spin Hall effect without spin-orbit coupling, offering an efficient route to spin current generation. We identify 23 candidate AFM1 materials by combining artificial intelligence, density functional theory (DFT), and symmetry analysis. Using a graph neural network model and incorporating AFM1-specific symmetry constraints, we screen Materials Project compounds for high-probability AFM1 candidates. DFT calculations show that AFM1 has the lowest energy among the tested magnetic configurations in 23 candidate materials. These include 3 experimentally verified AFM1 materials, 10 synthesized compounds with unknown magnetic structures, and 10 that are not yet synthesized.

# Introduction

Antiferromagnetic materials have gained significant attention in the context of next-generation spintronics due to their ultrafast spin dynamics, robustness against external magnetic fields, and the absence of stray magnetic fields [1]. Among them, parity-time-reversal-symmetric (PT-symmetric) odd-parity antiferromagnetic (AFM1) materials are particularly promising due to their unconventional symmetry properties and the novel physics they enable [2, 3]. These materials exhibit antiferromagnetic ordering that breaks space inversion (P) and time reversal (T) symmetries while preserving their combined PT symmetry.

A key consequence of this symmetry combination is the emergence of odd-parity energy contributions, $\varepsilon_{\text{odd}}(\mathbf{k})$, in the electronic dispersion. These terms lead to asymmetric band dispersion, enabling physical effects such as the magnetopiezoelectric effect, nonreciprocal conductivity, and photocurrent generation, making AFM1 materials promising platforms for unconventional quantum responses.

In addition to these effects, PT-symmetric magnets (AFM1 materials) have also been proposed as efficient spin current generators through a nonlinear spin Hall effect that does not rely on spin-orbit coupling [4]. This mechanism offers a symmetry-driven route for generating spin current in light-element materials, expanding the scope of spintronic applications beyond conventional SOC-based platforms.

Recent studies have further highlighted the fundamental significance of PT-symmetric antiferromagnets. For example, Matsyshyn *et al.* demonstrated that PT-symmetric magnets can host Kramers nonlinearity [5]. Meanwhile, Tang *et al.* showed that Dirac fermions can emerge in PT-symmetric antiferromagnetic systems [6]. These findings reinforce the relevance of AFM1 materials for exploring new quantum phases and guiding the development of future spintronic and electronic applications.

A systematic classification of AFM1 materials was carried out by Watanabe and Yanase, who identified 123 AFM1 materials through symmetry analysis of experimentally verified magnetic structures [2, 3]. However, discovering additional AFM1 materials remains crucial for uncovering new symmetry-driven physical phenomena and guiding experimental efforts. Since direct experimental determination of magnetic configurations is often resource-intensive, theoretical predictions play a critical role in accelerating the identification of candidate materials.

✉ wu67@uwm.edu

In this work, we present an artificial intelligence (AI)-assisted approach to predict AFM1 materials. We apply a graph neural network (GNN) model [7], incorporating AFM1-specific symmetry constraints in the dataset preparation, to screen Materials Project [8] compounds, followed by density functional theory (DFT) calculations to compare the energies of competing magnetic configurations. This combined approach yields 23 AFM1 candidate materials beyond the 123 materials identified by Watanabe and Yanase. These comprise 3 experimentally verified AFM1 materials, 10 synthesized materials with undetermined magnetic structures, and 10 computational predictions that have not yet been synthesized. These materials are further analyzed to determine their magnetic point groups (MPGs) and the symmetry-imposed forms of their odd-parity contributions, $\varepsilon_{\text{odd}}(\mathbf{k})$, along with the associated physical consequences, providing a foundation for future experimental and theoretical exploration of AFM1-driven physical phenomena.

# Results

## PT-Symmetric Magnets

Recently, there has been growing interest in a new class of magnetic materials called altermagnets, which exhibit novel physical properties arising from their symmetry [9, 10]. A recent wave of AI-assisted searches has been dedicated to identifying such materials [7]. In contrast, the focus of this study is on a different class of materials—parity-time-reversal-symmetric (PT-symmetric) magnets—where inversion symmetry plays a distinct role in shaping the magnetic structure.

Fig. 1 illustrates the key difference between altermagnets and PT-symmetric magnets by comparing the magnetic structures of $MnF_2$ [11], a representative altermagnet, and $FeGe_3$, a PT-symmetric magnet identified in this study. In $MnF_2$ (space group 136), the inversion centers of the nonmagnetic crystal coincide with the magnetic atom sites. In contrast, $FeGe_3$ (space group 194) realizes a PT-symmetric magnetic configuration, where the inversion centers lie between the magnetic atoms. The two magnetic sublattices are connected by the combined symmetry operation of spatial inversion (P) and time reversal (T). As a result, the magnetic order breaks both inversion and time-reversal symmetries individually, while preserving their product, PT.

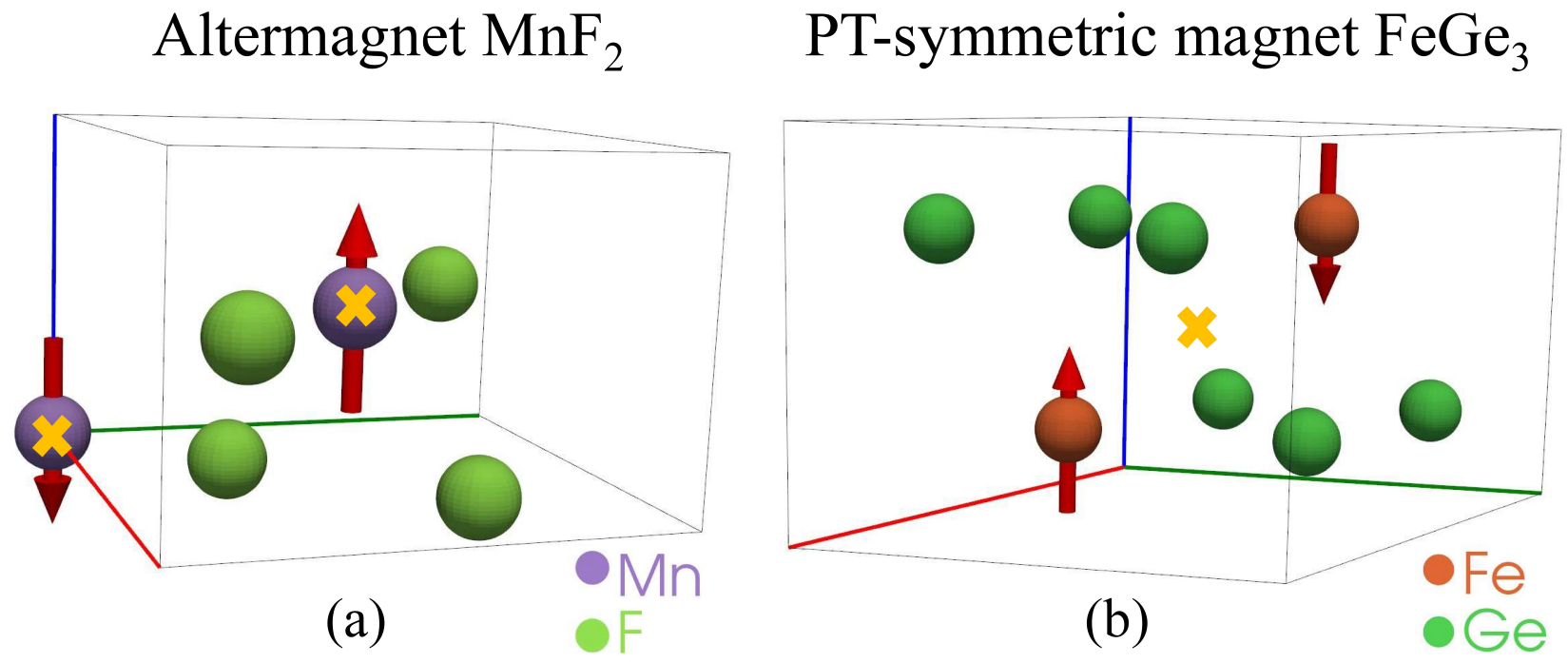

Figure 1: Comparison of magnetic structures in an altermagnet and a PT-symmetric magnet. The yellow crosses mark the inversion centers of the nonmagnetic crystal lattice. (a) $MnF_2$ (space group 136), a representative altermagnet, where the inversion centers coincide with the magnetic atom sites. (b) $FeGe_3$ (space group 194), a PT-symmetric antiferromagnet identified in this study, where the inversion centers lie between magnetic atoms, relating the two sublattices via combined inversion and time-reversal symmetry. Red arrows indicate the directions of magnetic moments.

This preservation of PT symmetry leads to distinctive physical consequences that differentiate PT-symmetric magnets from altermagnets and conventional antiferromagnets. The next two subsections explore two key classes of phenomena made possible by this symmetry structure.

## Novel Odd-Parity Energy Contributions and Their Impact

Materials with AFM1 order impose unique symmetry constraints on the electronic structure, giving rise to odd-parity energy contributions, denoted as $\varepsilon_{\text{odd}}(\mathbf{k})$. This term plays a fundamental role in modifying the energy dispersion, leading to unconventional emergent responses. In this subsection, we elucidate the origin of $\varepsilon_{\text{odd}}(\mathbf{k})$ through a tight-binding model for an AFM1 material and discuss its key physical consequences.

To understand how the AFM1 order gives rise to $\varepsilon_{\text{odd}}(\mathbf{k})$, we consider a symmetry-based tight-binding Hamiltonian for $TaFeO_4$ (Material No. 12 in Table 1). This material crystallizes in the monoclinic space group P2/c (13), with Fe atoms occupying the 2f Wyckoff positions, which lack local inversion symmetry. The unit cell contains two Fe atoms that are related by global inversion symmetry, forming two magnetic sublattices. The tight-binding model includes spin-independent hopping terms, SOC interactions, and an AFM1 order parameter. The Hamiltonian is expressed as

$$H = \varepsilon_{0,\mathbf{k}}\tau_0\sigma_0 + t_{x,\mathbf{k}}\tau_x\sigma_0 + t_{y,\mathbf{k}}\tau_y\sigma_0 + \lambda_{x,\mathbf{k}}\tau_z\sigma_x + \left(\lambda_{y,\mathbf{k}} + M_y\right)\tau_z\sigma_y + \lambda_{z,\mathbf{k}}\tau_z\sigma_z, \tag{1}$$

where $\tau_i$ are Pauli matrices encoding the Fe sublattice degree of freedom, and $\sigma_i$ are Pauli matrices encoding the electron spin degrees of freedom.

Here, $\varepsilon_{0,\mathbf{k}}$ represents the sublattice-independent dispersion term, while $t_{x,\mathbf{k}}$ and $t_{y,\mathbf{k}}$ describe inter-sublattice hopping terms that couple the two Fe sublattices. The SOC terms $\lambda_{i,\mathbf{k}}$ introduce spin-momentum coupling effects. Each coefficient in the Hamiltonian transforms according to irreducible representations of the $C_{2h}$ point group. The explicit momentum-dependent expressions for these coefficients are given by

$$\varepsilon_{0,\mathbf{k}} = \alpha\cos k_x + \beta\cos k_y + \gamma\cos k_z - \mu \tag{2a}$$

$$t_{x,\mathbf{k}} = t_x\cos\frac{k_z}{2} \tag{2b}$$

$$t_{y,\mathbf{k}} = t_{y1}\sin k_y\cos\frac{k_z}{2} + t_{y2}\sin k_x\sin k_y\sin\frac{k_z}{2} \tag{2c}$$

$$\lambda_{x,\mathbf{k}} = \lambda_{x1}\sin k_x + \lambda_{x2}\sin k_z \tag{2d}$$

$$\lambda_{y,\mathbf{k}} = \lambda_y\sin k_y \tag{2e}$$

$$\lambda_{z,\mathbf{k}} = \lambda_{z1}\sin k_x + \lambda_{z2}\sin k_z. \tag{2f}$$

The AFM1 order is introduced as an exchange field $M_y$ along the $y$-direction, represented by the term $M_y\tau_z\sigma_y$ in the Hamiltonian. The energy dispersion is obtained by treating $M_y$ as a perturbation, which leads to

$$E_\pm(\mathbf{k}) = \varepsilon_{0,\mathbf{k}} \pm \left(\sqrt{t_{x,\mathbf{k}}^2 + t_{y,\mathbf{k}}^2 + \lambda_{\mathbf{k}}^2} + \varepsilon_{\text{odd}}(\mathbf{k})\right), \tag{3}$$

where $\lambda_{\mathbf{k}}^2 = \lambda_{x,\mathbf{k}}^2 + \lambda_{y,\mathbf{k}}^2 + \lambda_{z,\mathbf{k}}^2$ and the odd-parity energy dispersion term is given by

$$\varepsilon_{\text{odd}}(\mathbf{k}) = \frac{M_y\lambda_y\sin k_y}{\sqrt{t_{x,\mathbf{k}}^2 + t_{y,\mathbf{k}}^2 + \lambda_{\mathbf{k}}^2}}. \tag{4}$$

In the absence of AFM1 order ($M_y = 0$), the energy dispersion remains symmetric under momentum inversion, satisfying $E(\mathbf{k}) = E(-\mathbf{k})$. However, when AFM1 order is present, the odd-parity nature of $\varepsilon_{\text{odd}}(\mathbf{k})$ introduces an asymmetry in the band structure, leading to $E(\mathbf{k}) \neq E(-\mathbf{k})$ (see Fig. 2). As a result, $\varepsilon_{\mathrm{odd}}(\mathbf{k})$ gives rise to a range of emergent responses, including the magnetopiezoelectric effect, nonreciprocal conductivity, and photocurrent generation, as identified in Ref. [2].

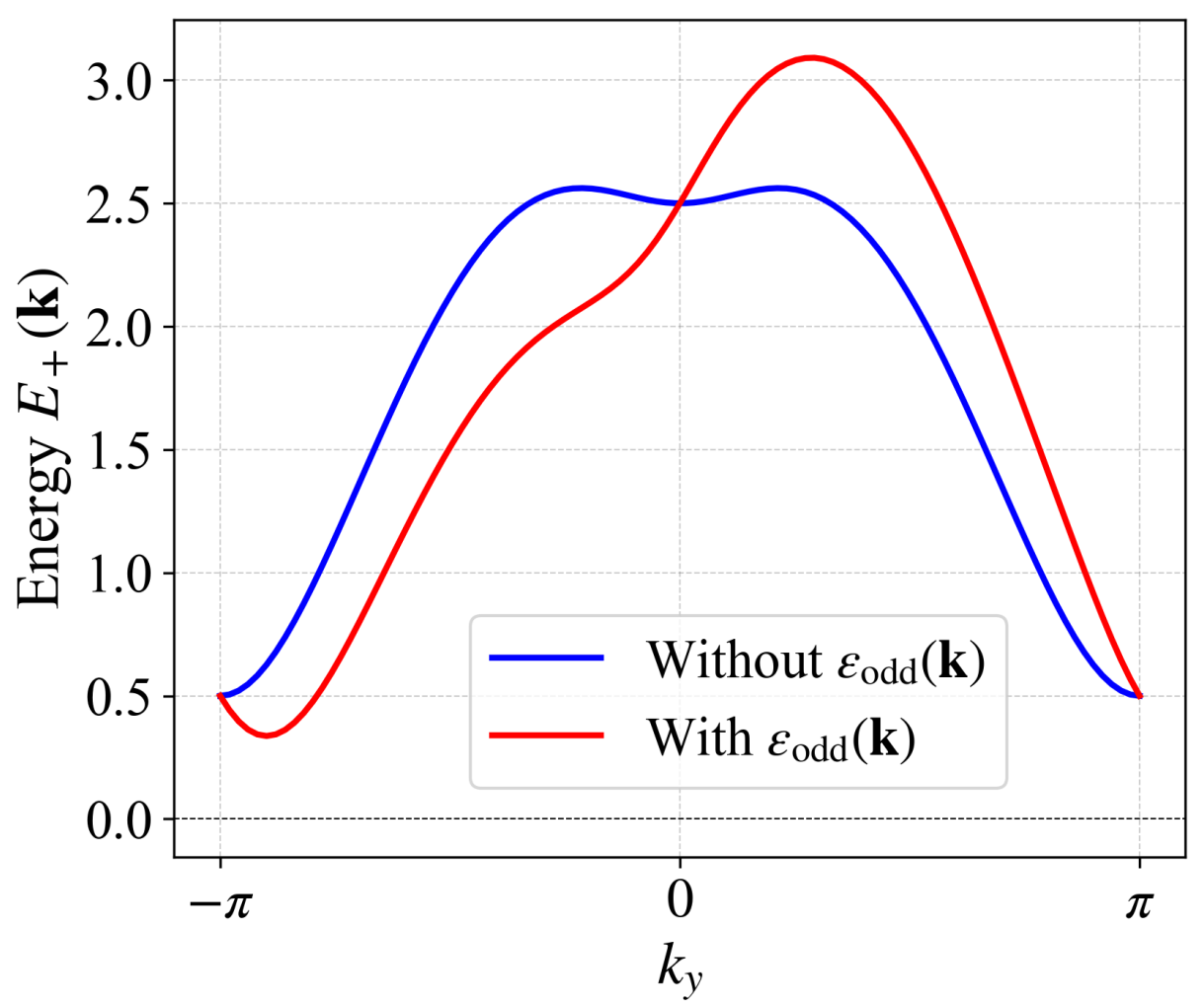

Figure 2: Energy dispersion $E_+(\mathbf{k})$ along the $k_y$-direction with fixed $k_x = 0$ and $k_z = 0$. The blue curve represents the dispersion without the odd-parity contribution $\varepsilon_{\mathrm{odd}}(\mathbf{k})$, while the red curve includes $\varepsilon_{\mathrm{odd}}(\mathbf{k})$. The asymmetry introduced by $\varepsilon_{\mathrm{odd}}(\mathbf{k})$ is evident, leading to a shift in the energy spectrum. The parameters used are $\alpha = 1.0$, $\beta = 1.0$, $\gamma = 1.0$, $\mu = 1.5$, $t_x = 1.0$, $t_{y1} = 0.8$, $t_{y2} = 0.5$, $\lambda_{x1} = 0.5$, $\lambda_{x2} = 0.3$, $\lambda_y = 1.0$, $\lambda_{z1} = 0.3$, $\lambda_{z2} = 0.2$, and $M_y = 1.0$.

The (inverse) magnetopiezoelectric (MPE) effect describes a lattice strain response induced by an applied electric current in metallic AFM1 materials [3, 12]. In contrast, in the conventional (inverse) piezoelectric effect, an applied electric field induces lattice strain in noncentrosymmetric materials. The underlying mechanism of the MPE effect is fundamentally different from conventional piezoelectricity. It originates from an asymmetric electronic band dispersion associated with an odd-parity energy contribution, $\varepsilon_{\mathrm{odd}}(\mathbf{k})$, which leads to an intrinsically momentum-asymmetric Fermi surface. In equilibrium, electron occupation remains balanced, and no net force acts on the lattice. However, when an electric current is applied, the Fermi surface shifts. This intrinsic asymmetry causes a nonuniform shift, leading to a momentum-dependent redistribution of conduction electrons. This redistribution modifies the electronic stress tensor, generating an effective electronic strain that couples to the lattice via electron-phonon interactions, ultimately resulting in a measurable structural deformation. The MPE effect has been experimentally observed in AFM1 materials such as $EuMnBi_2$ [13, 14] and $CaMn_2Bi_2$ [15], providing a mechanism for current-controlled lattice deformations with potential applications in antiferromagnetic spintronics and metallic electromechanical devices.

Beyond structural responses, asymmetric band dispersion in AFM1 materials also influences electronic transport. Nonreciprocal conductivity arises when an applied electric field induces an asymmetric current response. In these systems, $\varepsilon_{\mathrm{odd}}(\mathbf{k})$ modifies the velocity distribution of charge carriers, leading to a directional asymmetry in electrical transport. Despite extensive studies on nonreciprocal transport in T-symmetric materials, its realization in PT-symmetric AFM1 materials has received less attention and remains an important direction for future research. Such nonreciprocal transport effects could be useful for designing electronic components that allow current to flow more easily in one direction than the other, similar to a diode. This property may be applied in energy-efficient electronics and spintronic devices, where controlling the direction of current flow is crucial.

A related consequence of $\varepsilon_{\mathrm{odd}}(\mathbf{k})$ is photocurrent generation. In AFM1 materials, asymmetric band dispersion allows light irradiation to induce a direct current without an externally applied voltage. Unlike conventional photovoltaic effects that rely on built-in electric fields or externally applied voltage to drive charge separation, the photocurrent in these systems originates purely from band asymmetry. When light excites electrons to higher energy states, the asymmetric dispersion $\varepsilon_{\text{odd}}(\mathbf{k})$ causes an imbalance in carrier motion, leading to a net charge flow. In a symmetric band structure, photogenerated electrons and holes move equally in opposite directions, canceling any net current. However, in AFM1 materials, the asymmetric electronic band structure associated with $\varepsilon_{\text{odd}}(\mathbf{k})$ ensures that photoexcited carriers preferentially move in one direction, generating a measurable DC photocurrent. Like nonreciprocal conductivity and the magnetopiezoelectric effect, this phenomenon originates from $\varepsilon_{\text{odd}}(\mathbf{k})$. In particular, both photocurrent generation and nonreciprocal conductivity involve a directional charge transport response, with the former driven by light irradiation and the latter by an electric field. Photocurrent generation enabled by $\varepsilon_{\text{odd}}(\mathbf{k})$ suggests potential applications in optoelectronic devices and energy conversion technologies.

## Nonlinear Spin Hall Effect

Unlike the effects discussed in the preceding subsection, which rely on spin-orbit coupling to generate asymmetric band dispersion, the nonlinear spin Hall effect discussed here does not require SOC. Instead, it arises purely from the symmetry of the magnetic order in PT-symmetric magnets [4].

The tight-binding model introduced in the previous subsection can also host this SOC-free effect. In particular, the inter-sublattice hopping terms $t_{x,\mathbf{k}}$ and $t_{y,\mathbf{k}}$, which are spin-independent, play a key role in enabling a second-order spin transport response to an electric field. The result is a nonlinear spin Hall effect, where an applied electric field generates a transverse spin current.

This mechanism originates from the magnetic symmetry of the PT-symmetric AFM1 order. Since it does not rely on spin-orbit coupling, the nonlinear spin Hall effect can occur even in light-element compounds, offering a promising route for generating spin currents in antiferromagnetic materials.

## The 23 AFM1 candidate materials

Table 1 presents a summary of the 23 AFM1 candidate materials identified through the combined AI-based screening and DFT calculations described in the Methods section. For all 23 materials, the AFM1/AFM1a configuration was found to be the most stable magnetic state among the four magnetic configurations calculated, with a minimum energy difference of at least 1 meV/atom compared to the lowest non-AFM1/AFM1a configuration. These energy differences were determined using our converged DFT parameters, including a kinetic energy cutoff and a k-point grid specified in the Methods section. The materials are listed alongside their magnetic element, Materials Project ID, space group (SG), conductivity type, energy difference, experimental synthesis status, and reference sources.

The experimental status is classified into three categories: Verified AFM1, referring to materials that have been experimentally confirmed to exhibit AFM1 ordering; Synthesized, indicating materials that have been experimentally prepared but lack confirmed magnetic structure data; and Computational, representing materials identified through theoretical calculations without reported experimental synthesis, with the Materials Project database cited as the reference source. Among the 23 materials listed, 3 have been experimentally confirmed as AFM1 (the first two exhibiting collinear AFM1 order and the third noncollinear), 10 have been synthesized without magnetic verification, and the remaining 10 were computationally identified. This list serves as the foundation for further investigation into the physical properties of AFM1 materials.

Table 2 presents the MPGs and symmetry-derived forms of the odd-parity energy contributions $\varepsilon_{\text{odd}}(\mathbf{k})$ for the 23 AFM1 candidate materials introduced in Table 1. The term $\varepsilon_{\text{odd}}(\mathbf{k})$ is directly responsible for asymmetric band dispersion and underlies the three physical responses discussed. In addition, the MPG determines whether a material permits a nonlinear spin Hall effect in the absence of spin-orbit coupling, as identified by symmetry-based criteria developed in Ref. [4].

Table 1: Summary of the 23 AFM1 candidate materials identified through AI-based screening and DFT calculations. The columns present the material name, magnetic element, Materials Project (MP) ID, space group (SG), conductivity type, energy difference ($\Delta E$), experimental status, and reference source. Energy differences ($\Delta E$) are reported as the difference between the lowest energy of non-AFM1/AFM1a magnetic configurations (FM, AFM2, or AFM3) and the energy of AFM1/AFM1a configurations, measured in meV per atom. Larger $\Delta E$ values reflect stronger energetic stability of AFM1 over other configurations. Conductivity types are classified based on the band gap values provided by the Materials Project database. The experimental status is categorized as: Verified AFM1 (materials experimentally confirmed as AFM1 through neutron diffraction, where the first two listed materials in the table exhibit collinear AFM1 order and the third $NdB_4$ exhibits noncollinear AFM1 order), Synthesized (materials experimentally prepared with unconfirmed magnetic structures), and Computational (materials identified through theoretical calculations without reported synthesis). For computational materials, the Materials Project is cited as the reference source.

| No. | Material | Mag. | MP ID | SG | Conductivity | $\Delta E$ (meV/atom) | Status | Ref. |
|---|---|---|---|---|---|---|---|---|
| 1 | $BaMn_2Ge_2$ | Mn | mp-22412 | I4/mmm (139) | Metal | 7.94 | Verified AFM1 | [16] |
| 2 | $ErGe_3$ | Er | mp-513 | Cmcm (63) | Metal | -0.09 | Verified AFM1 | [17] |
| 3 | $NdB_4$ | Nd | mp-1632 | P4/mbm (127) | Metal | 2.01 | Verified AFM1 | [18] |
| 4 | $SmB_4$ | Sm | mp-8546 | P4/mbm (127) | Metal | 1.43 | Synthesized | [19] |
| 5 | $BaMn_2Sn_2$ | Mn | mp-22679 | I4/mmm (139) | Metal | 25.63 | Synthesized | [20] |
| 6 | $SrMn_2Ge_2$ | Mn | mp-21118 | I4/mmm (139) | Metal | 1.66 | Synthesized | [20] |
| 7 | $Ba_2Mn_2Bi_2O$ | Mn | mp-556391 | $P6_3/mmc$ (194) | Metal | 3.59 | Synthesized | [21] |
| 8 | $Ba_2Mn_2Sb_2O$ | Mn | mp-19213 | $P6_3/mmc$ (194) | Metal | 3.83 | Synthesized | [21–23] |
| 9 | $Ba_2Mn_2As_2O$ | Mn | mp-550454 | C2/m (12) | Metal | 23.66 | Synthesized | [22, 24] |
| 10 | $GdSn_2$ | Gd | mp-1071567 | Cmcm (63) | Metal | 1.07 | Synthesized | [25, 26] |
| 11 | GdSnGe | Gd | mp-1206580 | Cmcm (63) | Metal | 1.39 | Synthesized | [27, 28] |
| 12 | $TaFeO_4$ | Fe | mp-755628 | P2/c (13) | Semiconductor | 2.57 | Synthesized | [29] |
| 13 | $NaNiPO_4$ | Ni | mp-776294 | Pnma (62) | Semiconductor | 1.04 | Synthesized | [30] |
| 14 | $BaMn_2Bi_2$ | Mn | mp-1232615 | C2/m (12) | Metal | 6.56 | Computational | [8] |
| 15 | $BaMn_2As_2$ | Mn | mp-1232849 | C2/m (12) | Metal | 9.86 | Computational | [8] |
| 16 | $PmB_3$ | Pm | mp-862984 | $P6_3/mmc$ (194) | Metal | 2.03 | Computational | [8] |
| 17 | $FeGe_3$ | Fe | mp-1184472 | $P6_3/mmc$ (194) | Metal | 18.36 | Computational | [8] |
| 18 | $CoGe_2O_6$ | Co | mp-1353855 | $P2_1/c$ (14) | Semiconductor | 1.10 | Computational | [8] |
| 19 | $CaMn_2Sb_2$ | Mn | mp-1232722 | C2/m (12) | Semiconductor | 7.52 | Computational | [8] |
| 20 | $TaFeO_4$ | Fe | mp-755303 | C2/c (15) | Semiconductor | 3.42 | Computational | [8] |
| 21 | $TaMnO_4$ | Mn | mp-1208589 | P2/c (13) | Semiconductor | 1.90 | Computational | [8] |
| 22 | $NbCrO_4$ | Cr | mp-772660 | Pbcn (60) | Semiconductor | 6.84 | Computational | [8] |
| 23 | $Ho_4Zr_3O_{12}$ | Ho | mp-1224122 | $P\bar{1}$ (2) | Semiconductor | 9.38 | Computational | [8] |

Table 2: Summary of the magnetic point groups (MPGs) and symmetry-imposed odd-parity energy contributions $\varepsilon_{\text{odd}}(\mathbf{k})$ for the 23 AFM1 candidate materials. For each spin orientation along the Cartesian axes, the corresponding MPG is reported. The $\varepsilon_{\text{odd}}(\mathbf{k})$ terms are given as functions of the momentum components $k_i$. The last column illustrates magnetic structures with spin oriented along the z axis, which is the default configuration used in DFT calculations. The three experimentally verified AFM1 materials (Nos. 1–3) have magnetic structures determined by neutron diffraction and are available in the MAGNDATA database [31]. In these cases, the spin orientation is along the z axis (No. 1), within the $ab$-plane (No. 2), and noncollinear (No. 3), respectively.

| No. | Material | PG | Orient. | MPG | $\varepsilon_{\text{odd}}(\mathbf{k})$ | Mag. Structure (z) |
|---|---|---|---|---|---|---|
| 1 | $BaMn_2Ge_2$ | $D_{4h}$ | z | 4′/m′m′m | $\alpha k_x k_y k_z$ | 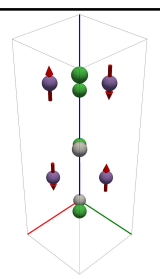 |
| 2 | $ErGe_3$ | $D_{2h}$ | $ab$-plane | 2/m′ | $\alpha_y k_y$ | 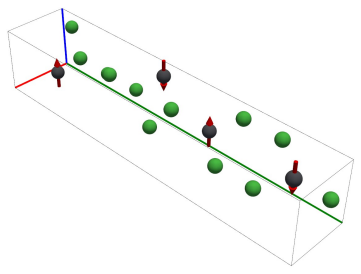 |
| 3 | $NdB_4$ | $D_{4h}$ | noncollinear | 2′/m | $\alpha_x k_x + \alpha_z k_z$ | 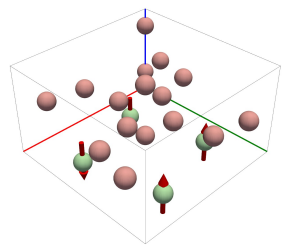 |
| 4 | $SmB_4$ | $D_{4h}$ | x | m′m′m′ | $\alpha k_x k_y k_z$ | 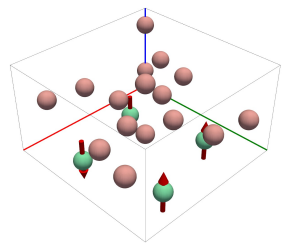 |
| | | | y, z | m′mm | $\alpha_z k_z$ | |
| 5 | $BaMn_2Sn_2$ | $D_{4h}$ | x, y | m′mm | $\alpha_z k_z$ | 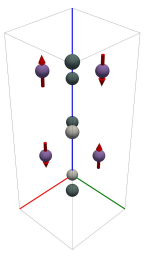 |
| | | | z | 4′/m′m′m | $\alpha k_x k_y k_z$ | |
| 6 | $SrMn_2Ge_2$ | $D_{4h}$ | x, y | m′mm | $\alpha_z k_z$ | 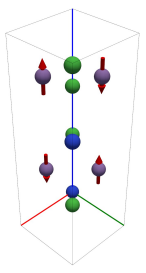 |
| | | | z | 4′/m′m′m | $\alpha k_x k_y k_z$ | |
| 7 | $Ba_2Mn_2Bi_2O$ | $D_{6h}$ | x, y | m′mm | $\alpha_z k_z$ | 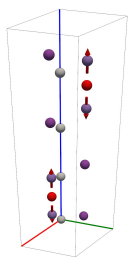 |
| | | | z | 6/m′m′m′ | $\alpha k_x k_y k_z \left(k_x^2 - 3k_y^2\right)\left(3k_x^2 - k_y^2\right)$ | |
| 8 | $Ba_2Mn_2Sb_2O$ | $D_{6h}$ | x, y | m′mm | $\alpha_z k_z$ | 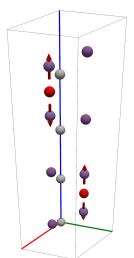 |
| | | | z | 6/m′m′m′ | $\alpha k_x k_y k_z \left(k_x^2 - 3k_y^2\right)\left(3k_x^2 - k_y^2\right)$ | |
| 9 | $Ba_2Mn_2As_2O$ | $C_{2h}$ | x, z | 2/m′ | $\alpha_y k_y$ | 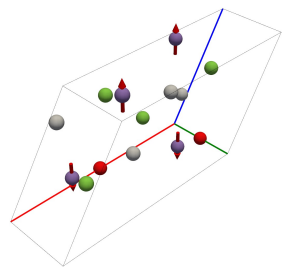 |
| | | | y | 2′/m | $\alpha_x k_x + \alpha_z k_z$ | |
| 10 | $GdSn_2$ | $D_{2h}$ | x, z | m′mm | $\alpha_z k_z$ | 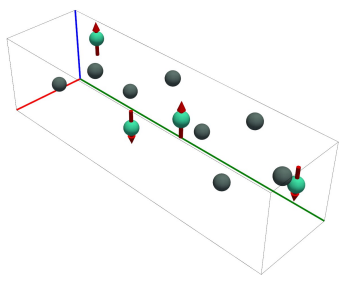 |
| | | | y | m′m′m′ | $\alpha k_x k_y k_z$ | |

TABLE 2. (Continued.)

| No. | Material | PG | Orient. | MPG | $\varepsilon_{odd}(\mathbf{k})$ | Mag. Structure (z) |
|---|---|---|---|---|---|---|
| 11 | GdSnGe | $D_{2h}$ | x, z | m′mm | $\alpha_z k_z$ | 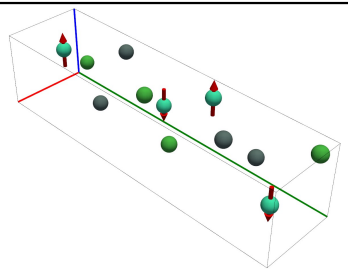 |
| | | | y | m′m′m′ | $\alpha k_x k_y k_z$ | |
| 12 | $TaFeO_4$ | $C_{2h}$ | x, z | 2′/m | $\alpha_x k_x + \alpha_z k_z$ | 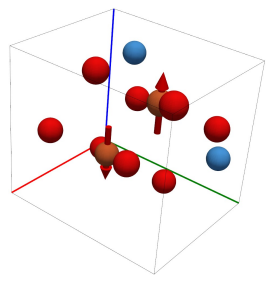 |
| | | | y | 2/m′ | $\alpha_y k_y$ | |
| 13 | $NaNiPO_4$ | $D_{2h}$ | x, y | m′mm | $\alpha_z k_z$ | 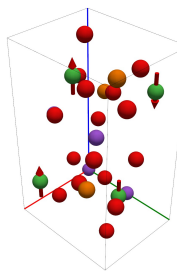 |
| | | | z | m′m′m′ | $\alpha k_x k_y k_z$ | |
| 14 | $BaMn_2Bi_2$ | $C_{2h}$ | x, z | 2/m′ | $\alpha_y k_y$ | 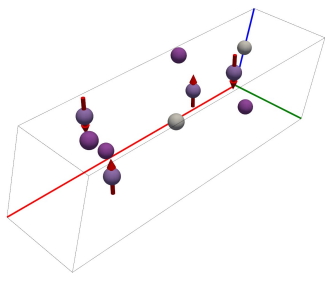 |
| | | | y | 2′/m | $\alpha_x k_x + \alpha_z k_z$ | |
| 15 | $BaMn_2As_2$ | $C_{2h}$ | x, z | 2/m′ | $\alpha_y k_y$ | 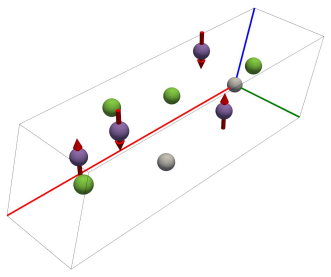 |
| | | | y | 2′/m | $\alpha_x k_x + \alpha_z k_z$ | |
| 16 | $PmB_3$ | $D_{6h}$ | x | m′mm | $\alpha_z k_z$ | 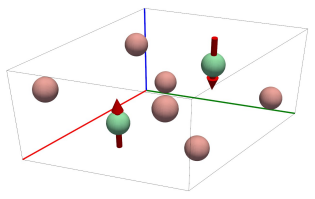 |
| | | | y | m′m′m′ | $\alpha k_x k_y k_z$ | |
| | | | z | 6′/mmm′ | $\alpha k_x \left(k_x^2 - 3k_y^2\right)$ | |
| 17 | $FeGe_3$ | $D_{6h}$ | x | m′mm | $\alpha_z k_z$ | 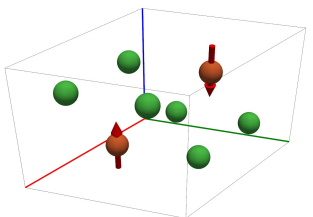 |
| | | | y | m′m′m′ | $\alpha k_x k_y k_z$ | |
| | | | z | 6′/mmm′ | $\alpha k_x \left(k_x^2 - 3k_y^2\right)$ | |
| 18 | $CoGe_2O_6$ | $C_{2h}$ | x, z | 2′/m | $\alpha_x k_x + \alpha_z k_z$ | 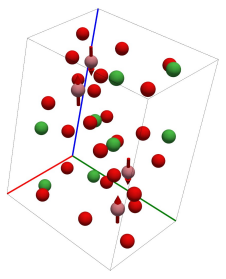 |
| | | | y | 2/m′ | $\alpha_y k_y$ | |
| 19 | $CaMn_2Sb_2$ | $C_{2h}$ | x, z | 2/m′ | $\alpha_y k_y$ | 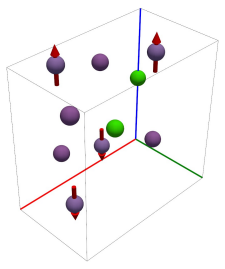 |
| | | | y | 2′/m | $\alpha_x k_x + \alpha_z k_z$ | |
| 20 | $TaFeO_4$ | $C_{2h}$ | x, z | 2′/m | $\alpha_x k_x + \alpha_z k_z$ | 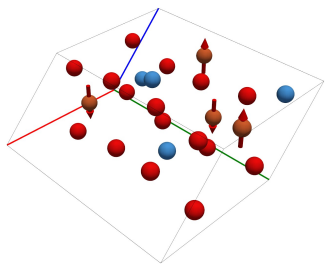 |
| | | | y | 2/m′ | $\alpha_y k_y$ | |
| 21 | $TaMnO_4$ | $C_{2h}$ | x, z | 2′/m | $\alpha_x k_x + \alpha_z k_z$ | 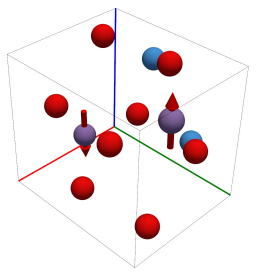 |
| | | | y | 2/m′ | $\alpha_y k_y$ | |

TABLE 2. (Continued.)

| No. | Material | PG | Orient. | MPG | $\varepsilon_{\mathrm{odd}}(\mathbf{k})$ | Mag. Structure (z) |
|---|---|---|---|---|---|---|
| 22 | $NbCrO_4$ | $D_{2h}$ | x, y | m′mm | $\alpha_z k_z$ | 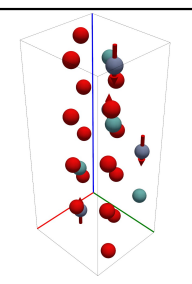 |
| | | | z | m′m′m′ | $\alpha k_x k_y k_z$ | |
| 23 | $Ho_4Zr_3O_{12}$ | $C_i$ | x, y, z | -1′ | $\alpha_x k_x + \alpha_y k_y + \alpha_z k_z$ | 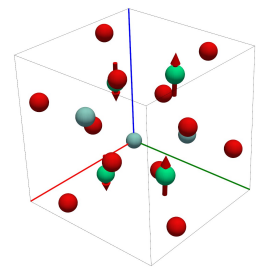 |

# Discussion

Watanabe and Yanase previously identified 123 AFM1 materials using a symmetry-based classification approach based on experimentally verified magnetic structures [2, 3]. Their work provided a foundational list of AFM1 materials, but to expand beyond this set and explore additional possibilities, we employed a different approach. Specifically, we conducted a large-scale screening of compounds from the Materials Project database, which contains both experimentally synthesized and computationally predicted compounds. To achieve this, we adopted the AI + DFT framework developed by Gao et al. [7]. Using their publicly available AI code, we trained a model for material screening and integrated it with first-principles DFT calculations to assess the energetic stability of candidate materials. This approach led to the identification of 23 promising AFM1 candidates: 3 experimentally verified AFM1 materials, 10 synthesized materials without confirmed magnetic structures, and 10 purely computational predictions. Further experimental studies, including synthesis and magnetic structure characterization, are necessary to fully validate these candidates.

Despite these promising results, limitations remain. First, our DFT calculations do not include SOC, which may affect the relative energy stability of magnetic configurations. Future studies incorporating SOC could refine these predictions. Second, while our AI-based screening significantly accelerates candidate identification, its performance is ultimately constrained by the quality of the training dataset, as well as the effectiveness of feature representations in capturing key material properties. Exploring alternative machine-learning architectures or integrating additional physical constraints could enhance predictive accuracy and improve the reliability of candidate selection.

To complement these computational efforts, experimental validation is crucial. We outline several key experimental directions that can further advance this research. First, synthesis efforts are needed for the 10 computationally predicted materials that have not yet been experimentally realized. Experimentalists can explore solid-state synthesis, thin-film deposition, or other growth techniques to fabricate these materials, enabling further magnetic characterization. Second, neutron diffraction and other magnetic structure characterization techniques are necessary to confirm the predicted AFM1 ordering in the identified materials. Additional methods such as muon spin rotation (μSR) and X-ray magnetic circular dichroism (XMCD) could provide complementary insights into their magnetic structures. Third, measurements of magnetopiezoelectric effects, nonreciprocal conductivity, and photocurrent generation are essential for verifying the physical consequences of $\varepsilon_{\mathrm{odd}}(\mathbf{k})$. Experimental studies on these transport and optical properties will help establish the functional significance of AFM1 materials. Finally, these AFM1 materials may have applications in antiferromagnetic spintronics, where their unique symmetry properties could enable efficient electrical control of antiferromagnetic order.

Beyond the specific focus on AFM1 materials, the methodological framework developed in this study is broadly applicable. The combination of AI-based screening and DFT calculations provides a scalable framework for discovering materials with targeted magnetic and electronic properties. This framework was originally proposed by Gao et al. [7] for identifying altermagnets. While our study focuses on AFM1 materials, the same methodology can be extended to identify other quantum materials exhibiting unconventional electronic, magnetic, or topological behaviors. By adjusting the training dataset and selection criteria, this approach could systematically explore materials with targeted symmetry, electronic, and magnetic properties, offering a powerful tool for accelerating quantum materials discovery.

Beyond their intrinsic properties, AFM1 materials could also serve as precursors for realizing altermagnetic behavior under controlled symmetry-breaking conditions. In their native AFM1 state, these materials exhibit PT symmetry, ensuring that their band structures remain degenerate despite the breaking of inversion symmetry (P) by magnetic order. This PT-protected degeneracy is a fundamental property of these materials and persists across the entire Brillouin zone. However, if the crystal structure itself also lacks inversion symmetry—due to strain, substrate engineering, or chemical modification—then PT symmetry is no longer preserved, leading to momentum-dependent spin splitting, a key characteristic of altermagnets [9, 10]. This mechanism has been discussed recently in Mazin et al. [32], where PT-invariant AFM materials were shown to exhibit altermagnetic spin textures upon structural inversion breaking. For instance, in FeSe, introducing a substrate resulted in altermagnetic-like band splitting, demonstrating a viable route for generating altermagnetic states from PT-invariant AFM systems. Given this, the AFM1 materials identified in this study could serve as potential precursors to altermagnets, provided that appropriate structural modifications are introduced. This suggests a natural route toward realizing altermagnetic properties in experimentally accessible systems. Future work could explore whether selective substrate interactions, strain engineering, or layer-dependent chemical modifications can effectively induce altermagnetism in these materials, potentially expanding their functional applications in next-generation spintronic devices. These possibilities further highlight the relevance of our findings in broader quantum material contexts.

In summary, we applied an AI-assisted screening approach, combined with first-principles DFT calculations, to identify AFM1 materials. This study demonstrates the effectiveness of AI-assisted screening for targeted material discovery, expanding the search for AFM1 candidates beyond previously identified materials. This approach identified 23 AFM1 candidate materials, including 3 that have been experimentally verified as AFM1, 10 that have been synthesized but lack confirmed magnetic structures, and 10 that remain purely computational predictions. This study highlights $\varepsilon_{\rm odd}(\mathbf{k})$ as an intrinsic consequence of the symmetry properties of AFM1 materials and its essential role in emergent quantum phenomena, including the magnetopiezoelectric effect, nonreciprocal conductivity, and photocurrent generation. In addition, PT-symmetric AFM1 materials enable a nonlinear spin Hall effect that does not require spin-orbit coupling, offering a symmetry-based route for spin current generation in antiferromagnetic systems.

While this study applies a systematic AI + DFT framework for AFM1 material discovery, further refinements are necessary. Experimental validation remains crucial, particularly for confirming the AFM1 ordering in synthesized materials and achieving the first synthesis of the computationally predicted candidates. Additionally, refining AI-based screening strategies and expanding material databases may further improve the identification of promising AFM1 materials.

Our results expand the search for AFM1 materials, identifying promising candidates for further theoretical and experimental exploration and advancing the understanding of symmetry-driven quantum phenomena. These materials could be of interest for applications in antiferromagnetic spintronics, where their unique symmetry properties provide new opportunities for efficient electrical control of magnetic order.

# Methods

## AI-Based Screening

To identify AFM1 materials, we applied the graph neural network (GNN) framework developed by Gao et al.[7], originally used for the discovery of altermagnetic materials. The model architecture and hyperparameters were adopted without modification to ensure consistency with their validated framework, which demonstrated strong performance on a closely related classification task. The primary difference in our study is the application to AFM1 materials, with only minimal differences in dataset preparation compared to Gao et al.'s screening for altermagnetic materials. For full technical details of the model architecture, hyperparameters, and publicly available code, readers are referred to Gao et al.'s original work. The workflow, including dataset preparation and model application for AFM1 screening, is outlined below.

To train the GNN model and screen materials for AFM1 characteristics, three datasets were prepared: positive samples, negative samples, and candidate materials. These datasets were constructed using Crystallographic Information Files (CIFs) obtained from the Materials Project database [8]. The CIF files contain the crystal structure information of these materials, providing the foundation for constructing structural representations used in the GNN model. Positive samples correspond to experimentally confirmed AFM1 materials, negative samples are materials that cannot host AFM1 states based on symmetry analysis, and candidate materials constitute a large dataset from which the promising AFM1 candidate materials were identified.

The positive samples were obtained from the 123 experimentally confirmed AFM1 materials identified by Yanase et al. [2, 3]. CIF files for 104 of these materials were available in the Materials Project database. Of these, 90 materials were used to construct the positive sample dataset for model training, while the remaining 14 positive samples were initially reserved as part of a test dataset for model evaluation, following standard machine learning practices. To maximize data utilization, these 14 positive samples were later incorporated into a second round of fine-tuning, as described later in this section.

For negative samples and candidate materials, the corresponding CIF files were obtained from the Materials Project database following a systematic filtering procedure. At the time of our data collection, the database contained a total of 154,718 materials. While the database is continuously updated, the selection criteria, data extraction procedure, and workflow, including the code implementation, remain consistent and repeatable. To construct our datasets, we considered materials containing magnetic elements, focusing on compounds with 3d transition metals or 4f rare-earth elements, as described in Ref. [7]. For simplicity, we limited the selection to materials with a single type of magnetic atom. The selected materials were then divided into two groups based on whether the primitive unit cell contains an odd or even number of magnetic atoms. Materials with an odd number of magnetic atoms in the primitive unit cell were directly included in the negative sample dataset. Such configurations are generally inconsistent with the symmetry constraints required for AFM1 ordering. For materials with an even number of magnetic atoms, an additional symmetry-based filtering step was applied using the following three structural conditions: (1) the crystal possesses inversion symmetry, (2) the primitive unit cell contains exactly two magnetic atoms, and (3) both magnetic atoms coincide with inversion centers within the primitive unit cell. Materials satisfying all three conditions were deemed structurally incompatible with AFM1 ordering and were therefore included in the negative sample set. Materials with an even number of magnetic atoms that did not satisfy all three conditions were retained as candidate materials for AFM1 discovery, as they remained structurally consistent with the possibility of hosting AFM1 ordering. The negative sample dataset consisted of the two parts described above. A total of 83 negative samples were set aside, along with the 14 positive samples previously reserved, to form the test dataset used for model evaluation. However, as the primary goal of this study is material discovery and exploration of physical properties rather than model validation, results from this test set will not be further discussed. After excluding the 83 reserved negative samples, a total of 21,600 negative samples were obtained. After excluding 69 materials already present in Refs. [2, 3], the final candidate material dataset consisted of 45,212 materials. The three prepared datasets were then used to train the GNN model and screen for AFM1 materials, as described below.

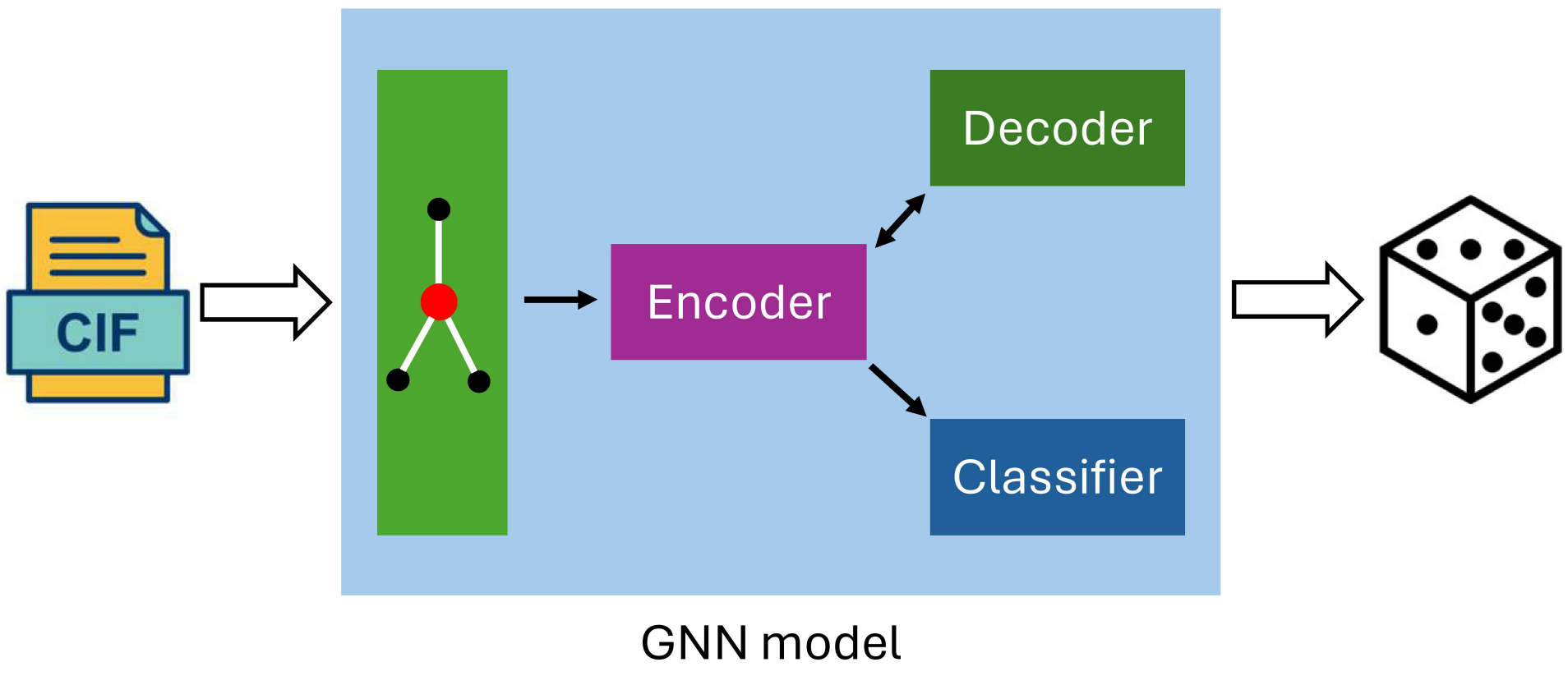

Figure 3: Schematic representation of the graph neural network (GNN) framework used for AFM1 material screening. The model consists of four main components: (1) a crystal graph conversion module that transforms CIF files into graph representations, (2) an encoder that extracts structural features and generates an encoded feature vector, (3) a decoder used during pre-training to reconstruct the original graph, and (4) a classifier that assigns a probability score for AFM1 ordering. The workflow involves three stages: pre-training, fine-tuning, and prediction, where different components are activated accordingly.

The GNN model employed in this study, developed by Gao et al. [7], takes CIF files as input and outputs a probability score indicating the likelihood of AFM1 ordering. It consists of four primary components: a crystal graph conversion module, an encoder, a decoder, and a classifier, which operate across three workflow stages—pre-training, fine-tuning, and prediction—with distinct roles described below. The crystal graph conversion module, active at all stages, converts CIF files into crystal graph representations where nodes correspond to atoms and edges represent structural connections between atoms, making the data suitable for the GNN framework. The encoder extracts abstract structural features from the crystal graph of each material and generates an encoded feature vector space, where each material is represented as a point described by a numerical feature vector. This feature space serves as the input for both the decoder and the classifier. During pre-training, the encoder and decoder work together: the encoder learns structural patterns from a large, unlabeled dataset, while the decoder attempts to reconstruct the original crystal graph to guide feature learning by minimizing a reconstruction loss function. In the subsequent fine-tuning and prediction stages, the decoder is inactive. The classifier uses the encoded features to generate a probability score for AFM1 ordering, with fine-tuning specifically refining the classifier by minimizing a classification loss function using a labeled dataset, where materials are explicitly categorized as either AFM1 or non-AFM1. A schematic representation of the GNN architecture is shown in Fig. 3.

In the pre-training stage, both the encoder and decoder are in use. The encoder processes crystal graph representations generated from a large combined dataset of 45,212 candidate materials and 21,600 negative samples, extracting abstract structural features and generating an encoded feature vector that serves as a compressed structural descriptor of each material. The encoded feature vector is directly determined by the encoder's model parameters, which are iteratively updated during pre-training to better capture structural patterns. The decoder, while actively reconstructing the original crystal graph from the encoded feature vector, does not involve trainable parameters but assists in guiding the encoder's feature learning. The reconstruction loss function, which measures the difference between the reconstructed and original graphs, is minimized during pre-training. This loss provides feedback that drives the encoder's parameter updates, ensuring that the structural features extracted by the encoder become more meaningful and representative of the original crystal structure. The purpose of pre-training is to allow the encoder to learn a general structural representation of materials from a large, unlabeled dataset, ensuring the extracted features capture meaningful structural patterns. The learning process and convergence behavior during pre-training are illustrated in the learning curve provided in the Supplementary Material.

In the fine-tuning stage, the encoder continues to function as a feature extractor, and its model parameters are further updated alongside the classifier. The encoded feature vectors generated by the pre-trained encoder are passed to the classifier, which consists of a series of fully connected layers designed to map the extracted features to a classification score. Fine-tuning was performed using 90 positive samples and 21,600 negative samples specific to AFM1 materials. The classifier was fine-tuned on this labeled data, and a classification loss function, which measures the difference between the predicted and true material labels, was minimized to refine the classifier's model parameters, improving its ability to distinguish between AFM1 and non-AFM1 materials. Unlike pre-training, where the focus is on learning general structural features from a large, unlabeled dataset, fine-tuning specifically adapts the model to classify AFM1 materials using labeled data.

In the prediction stage, the pre-trained encoder and fine-tuned classifier were used to evaluate 45,212 candidate materials. The encoder extracted structural features from the crystal graph representations, and the classifier generated a probability score indicating the likelihood of AFM1 ordering. No parameter updates occurred during this phase, as the model simply applied the learned model parameters from the fine-tuning stage to classify the materials. The fully trained GNN model identified 276 candidate materials with a classification score exceeding 0.9. These top candidates were subsequently subjected to DFT calculations for further validation of their AFM1 ordering potential.

## DFT Calculations

DFT calculations were performed to further refine the screening results obtained from the GNN model and verify the stability of AFM1 ordering in the selected candidate materials. We employed the Quantum ESPRESSO (QE) software package [33, 34], a plane-wave DFT code, in combination with the Atomic Simulation Environment (ASE) [35], a Python library that interfaces with QE to streamline the construction and simulation of magnetic configurations. The Standard Solid-State Pseudopotentials (SSSP) PBE Efficiency v1.3.0 library [36] was utilized for all calculations.

Out of the 276 materials identified with a classification score exceeding 0.9 from the GNN screening, 124 centrosymmetric materials containing 2 or 4 magnetic atoms in the unit cell were selected for DFT calculations, focusing on simpler structures to reduce computational cost. Four magnetic configurations were considered for each material: ferromagnetic (FM) and three antiferromagnetic (AFM) configurations, labeled AFM1, AFM2, and AFM3, representing the three possible collinear AFM arrangements for a material with four magnetic atoms in the unit cell (see Fig. 4). We define AFM1 as the PT-symmetric odd-parity antiferromagnetic configuration and construct it first for each material where this spin arrangement is achievable. The other configurations, AFM2 and AFM3, are generated algorithmically as contrasting magnetic states for energy comparisons, with their labels assigned according to the sequence in which the code generates distinct spin arrangements rather than predefined configurations. In some cases, a material may exhibit a variation of AFM1, denoted as AFM1a, where the magnetic moments satisfy the AFM1 definition but differ in their specific spin arrangements. In such instances, the three AFM configurations are labeled in order as AFM1, AFM1a, and AFM2, with AFM3 not present. For materials with two magnetic atoms in the unit cell where AFM1 is allowed, a supercell is constructed by doubling the unit cell along the z-axis to generate contrasting magnetic configurations, allowing AFM2 and AFM3 to be constructed.

For the 124 selected materials, those confirmed to be unable to host an AFM1 structure were excluded from further calculations. DFT calculations were performed on the remaining materials using a kinetic energy cutoff of 100 Ry and a $10 \times 10 \times 10$ $\mathbf{k}$-point grid. The total energies of all four magnetic configurations were computed and compared for each material. Materials where the AFM1 configuration exhibited the lowest total energy, with a relative energy difference exceeding 1 meV/atom compared to the other configurations, were selected for further analysis. After this screening, 19 materials satisfied this energy criterion.

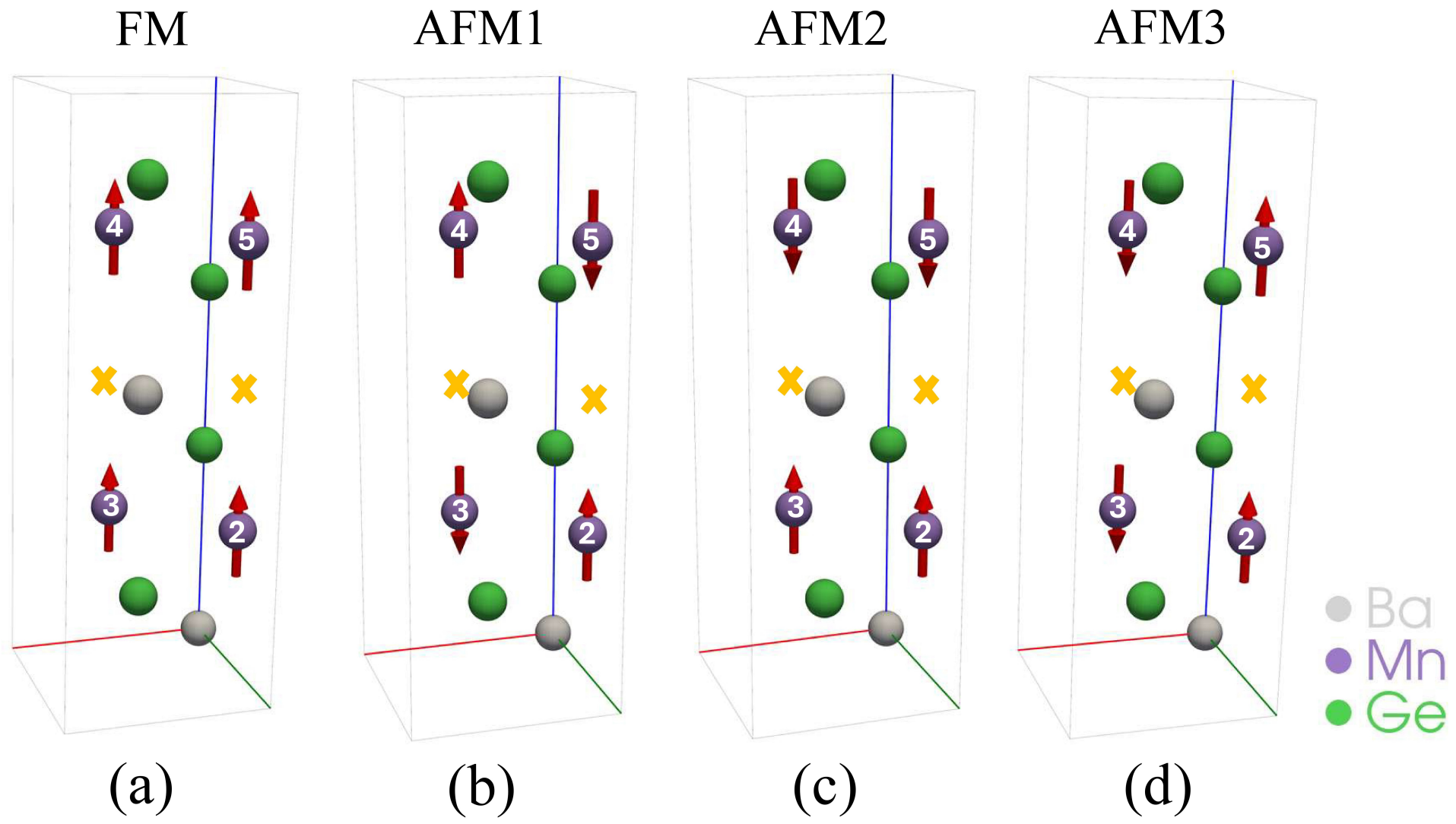

Figure 4: Magnetic configurations considered in this study for $BaMn_2Ge_2$ (mp-22412), which crystallizes in the tetragonal space group I4/mmm (139). It is one of the three experimentally verified AFM1 materials in our candidate list. The crystal structure consists of Ba (gray, Wyckoff position 2a), Mn (purple, 4d), and Ge (green, 4e). Red arrows indicate the spin orientations of Mn atoms. The yellow crosses indicate inversion centers located between the two Mn atoms in each inversion-related pair: (Mn2, Mn5) and (Mn3, Mn4). Additionally, centering translations relate the Mn pairs (Mn2, Mn4) and (Mn3, Mn5) with the centering vector (0.5, 0.5, 0.5). Subfigures (a)–(d) show the following magnetic configurations, respectively: (a) FM (ferromagnetic, all spins aligned), (b) AFM1 (the PT-symmetric odd-parity antiferromagnetic state), (c) AFM2 (an alternative antiferromagnetic state), (d) AFM3 (another competing antiferromagnetic configuration).

To maximize the discovery of promising AFM1 materials using the GNN model, the 19 identified materials and the 14 positive samples previously reserved for testing were added to the positive sample dataset. A second fine-tuning of the GNN model was performed using this expanded positive dataset alongside the original negative dataset. Candidate materials with classification scores exceeding 0.9 were then subjected to DFT calculations. Five additional materials were identified where AFM1 had the lowest energy, again with a relative energy difference greater than 1 meV/atom. These five materials were added to the positive sample set for a third round of fine-tuning and subsequent DFT calculations, yielding one more AFM1 candidate. To expand the search further, materials with classification scores between 0.5 and 0.9 from the combined results of the three rounds of fine-tuning were also examined. After applying the same selection criteria (centrosymmetric structures with 2 or 4 magnetic atoms, DFT energy comparisons, and literature verification), one additional material was identified with a high energy difference of 18 meV/atom between AFM1 and other configurations. In total, 26 materials were identified where the AFM1 configuration was the most stable among the four tested magnetic states. To further ensure the robustness of these results, two additional steps were performed: a literature search and convergence testing.

Literature searches were conducted using databases such as the Materials Project [8], MAGNDATA [31], the Crystallography Open Database (COD) [37], and the Materials Platform for Data Science (MPDS) [38], as well as general literature search engines, to verify the synthesis status and magnetic ordering for these materials. Two materials, $Mn_2TeO_6$ (mp-1210598) [39] and ErGe (mp-2264) [40], were determined by neutron diffraction to have magnetic structures that are not AFM1. After excluding these two materials based on the literature search results, a total of 24 materials remained on the list.

To ensure numerical accuracy, further convergence tests were performed on this set. The kinetic energy cutoff and **k**-point grid were tested using the FM configuration. Using the converged parameters, the magnetic configurations were recalculated, and the relative energy rankings among configurations remained consistent for all materials except for $ErGe_3$ (mp-513) and $PrB_4$ (mp-12569). For mp-513, the FM configuration was found to be 0.09 meV/atom lower in energy than AFM1, yet the material was retained due to experimental confirmation of its AFM1 ordering. For $PrB_4$, the AFM2 configuration was identified as the lowest energy state, with an energy difference of 0.8573 meV/atom below AFM1a, leading to its exclusion from the final AFM1 candidate list. By excluding $PrB_4$, the final shortlist was reduced to 23 promising AFM1 candidate materials.

Among the 23 AFM1 candidate materials, three were experimentally confirmed as AFM1 (two collinear and one noncollinear), 10 were experimentally synthesized but lacked confirmed magnetic ordering, and 10 had no reported experimental synthesis. The Results section below provides further details on these materials, including their classification and key properties, while the Supplementary Material presents the energy comparisons of different magnetic configurations and visualizations of their structures.

# Data availability

CIF files used in this study were obtained from the Materials Project database (https://materialsproject.org). Part of the data generated during this work, including AI screening outputs, DFT results, and structure visualizations, is provided in the Supplementary Material. Additional data are available from the corresponding author upon reasonable request.

# Code availability

The AI model used in this study is based on the publicly available implementation by Gao et al., available at https://github.com/zfgao66/MatAltMag, with minor modifications to the negative dataset preparation. Custom Python codes developed by H.W. for the DFT workflow (including automation of input generation and job submission via the Atomic Simulation Environment (ASE) and Quantum ESPRESSO), as well as for constructing and visualizing magnetic structures, are available from the corresponding author upon reasonable request.

# Acknowledgements

We acknowledge the CyberInfrastructure Comprehensive, Applied and Tangible Summer School (CIber-CATSS), hosted by Philip Chang, for providing essential training in machine learning and data science methodologies used in this study. We also thank the High Performance Computing (HPC) Service at the University of Wisconsin-Milwaukee for providing computational resources for the AI-based screening and density functional theory calculations. We are grateful to Ze-Feng Gao, Philip Chang, Michael Weinert, and Tatsuya Shishidou for their insightful discussions throughout this project. This work was supported by the National Science Foundation (NSF) through the Designing Materials to Revolutionize and Engineer our Future (DMREF) program under Grant No. 144-4-AAM6666, Discovery of Novel Magnetic Materials through Pseudospin Control.

# Author contributions

D.F.A. conceived and designed the project, supervised the research, and provided guidance on the theoretical analysis. H.W. carried out the AI-based screening and DFT calculations, analyzed the magnetic symmetry and physical implications, and wrote the initial draft of the manuscript. Both authors revised the manuscript.

# Competing interests

The authors declare no competing interests.

# Additional information

## Supplementary information

Supplementary material is provided as a separate file.

# Supplementary Material for AI-predicted PT-symmetric magnets

Hao Wu and Daniel F. Agterberg

Department of Physics, University of Wisconsin–Milwaukee, Milwaukee, Wisconsin 53201, USA

# Contents

# Learning Curve of the GNN Model During Pre-Training

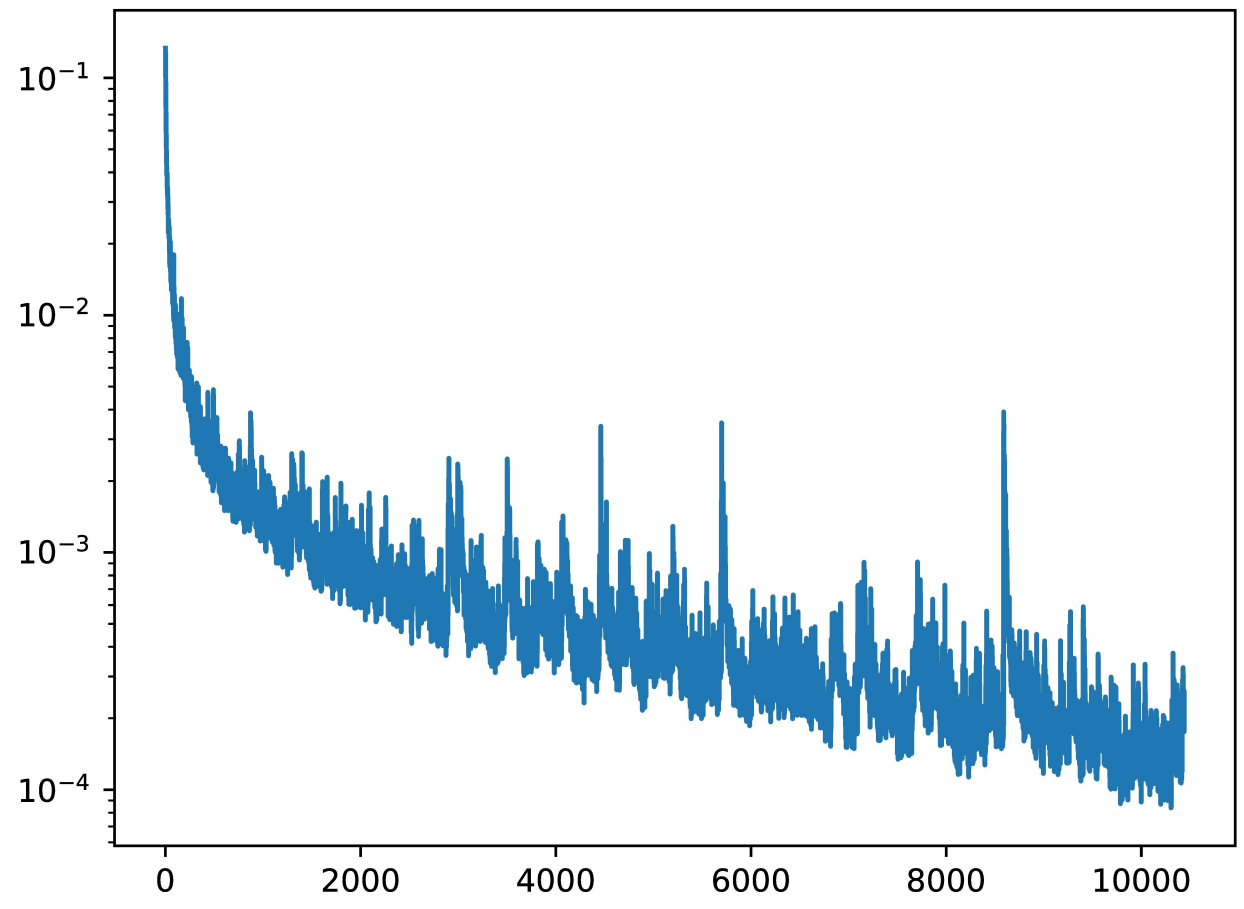

Figure S1: Learning curve of the GNN model during the pre-training stage, showing the total loss $\mathcal{L}$ as a function of iterations. The total loss is a weighted sum of the node feature reconstruction loss ($\mathcal{L}_s$) and the neighborhood reconstruction loss ($\mathcal{L}_p$), where $\mathcal{L}_s$ is based on Mean Squared Error (MSE) for node features, and $\mathcal{L}_p$ is computed using the 2-Wasserstein distance for neighborhood reconstruction. The x-axis represents the number of iterations, where each iteration corresponds to an update step processing a batch of 64 crystal structures. The model was trained with a learning rate of $1 \times 10^{-3}$, following the setup of Ze-Feng Gao et al., AI-accelerated discovery of altermagnetic materials, National Science Review, 12(4):nwaf066 (2025). The model was trained with a learning rate of $1 \times 10^{-3}$, following the setup of Ze-Feng Gao et al., *AI-accelerated discovery of altermagnetic materials*, *National Science Review*, 12(4):nwaf066 (2025), https://doi.org/10.1093/nsr/nwaf066. The consistent downward trend of the loss function indicates successful optimization in the pre-training stage, demonstrating the model's effective learning of structural representations.

# Magnetic Configuration Visualizations

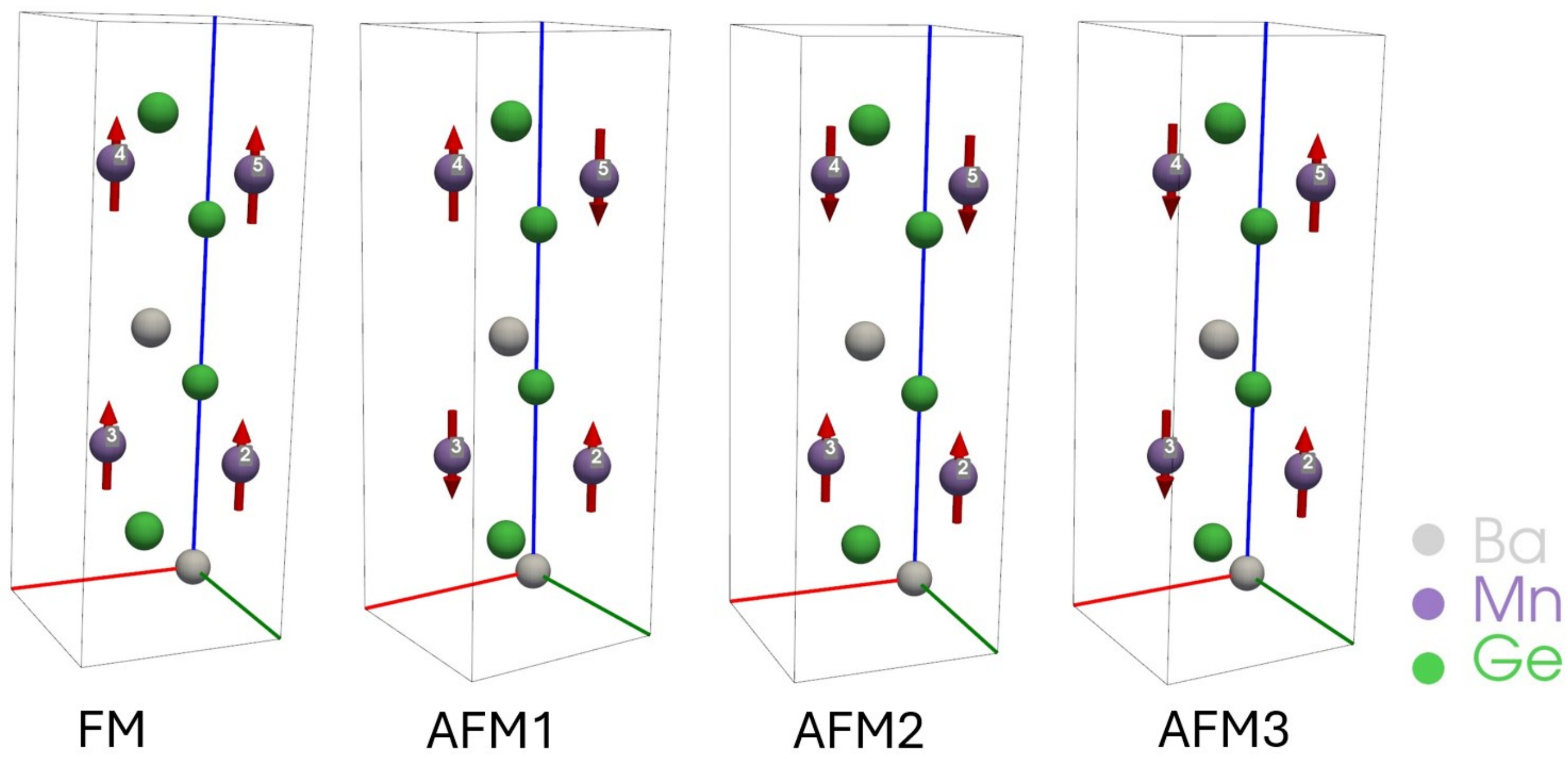

Figure S2: Magnetic configurations of material No. 1, $BaMn_2Ge_2$ (mp-22412), with space group I4/mmm (139). Wyckoff positions: Ba (2a), Mn (4d), Ge (4e). Inversion-related Mn pairs: (Mn2, Mn5) and (Mn3, Mn4). Centering-translation-related Mn pairs: (Mn2, Mn4) and (Mn3, Mn5), with the centering vector (0.5, 0.5, 0.5).

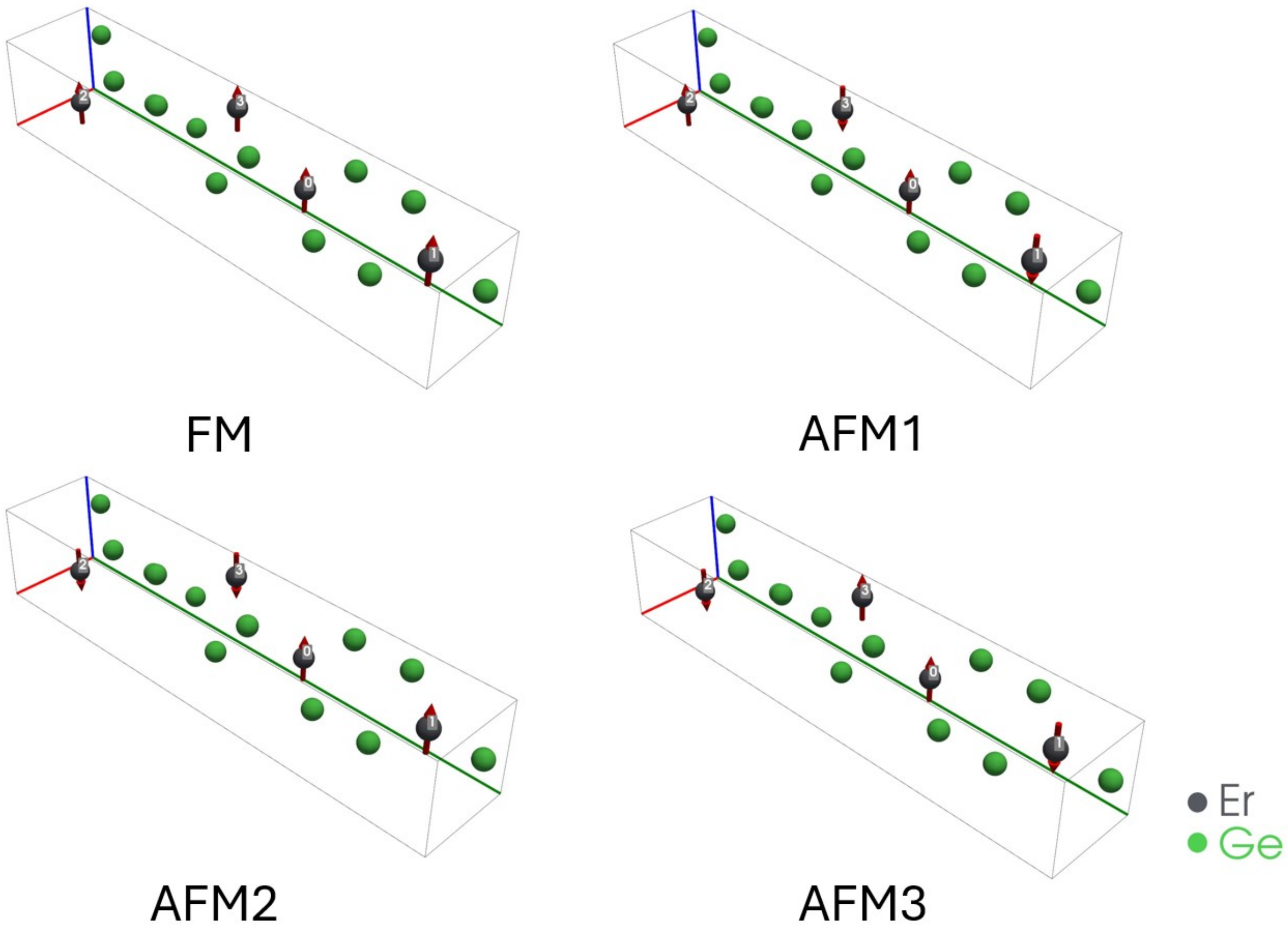

Figure S3: Magnetic configurations of material No. 2, $ErGe_3$ (mp-513), with space group Cmcm (63). Wyckoff positions: Er (4c), Ge (4c, 4c, 4c). Inversion-related Er pairs: (Er0, Er3) and (Er1, Er2). Centering-translation-related Er pairs: (Er0, Er2) and (Er1, Er3), with the centering vector (0.5, 0.5, 0).

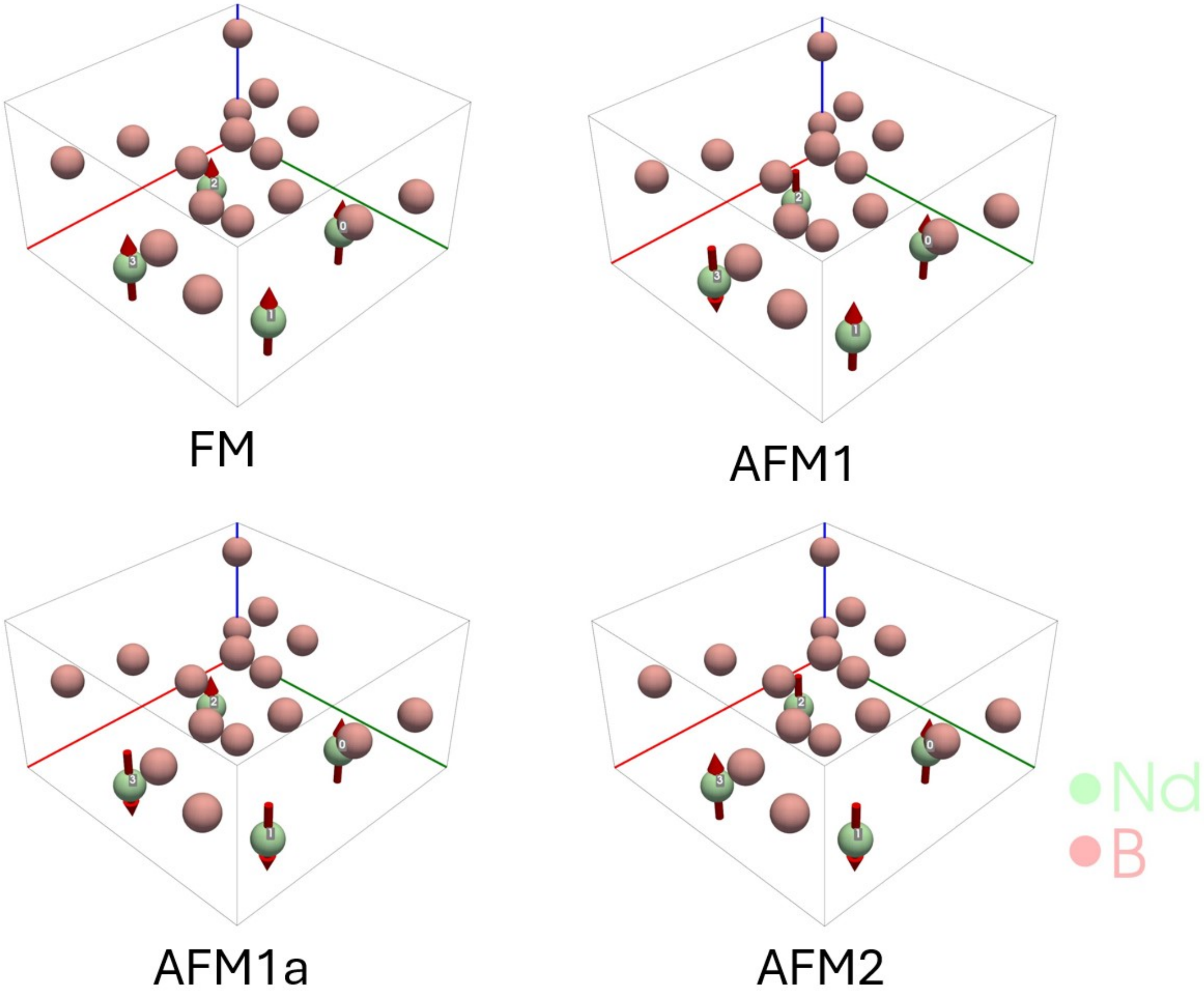

Figure S4: Magnetic configurations of material No. 3, $NdB_4$ (mp-1632), with space group P4/mbm (127). Wyckoff positions: Nd (4g), B (4h, 8j, 4e). Inversion-related Nd pairs: (Nd0, Nd3) and (Nd1, Nd2). No Nd atoms are related by centering translations.

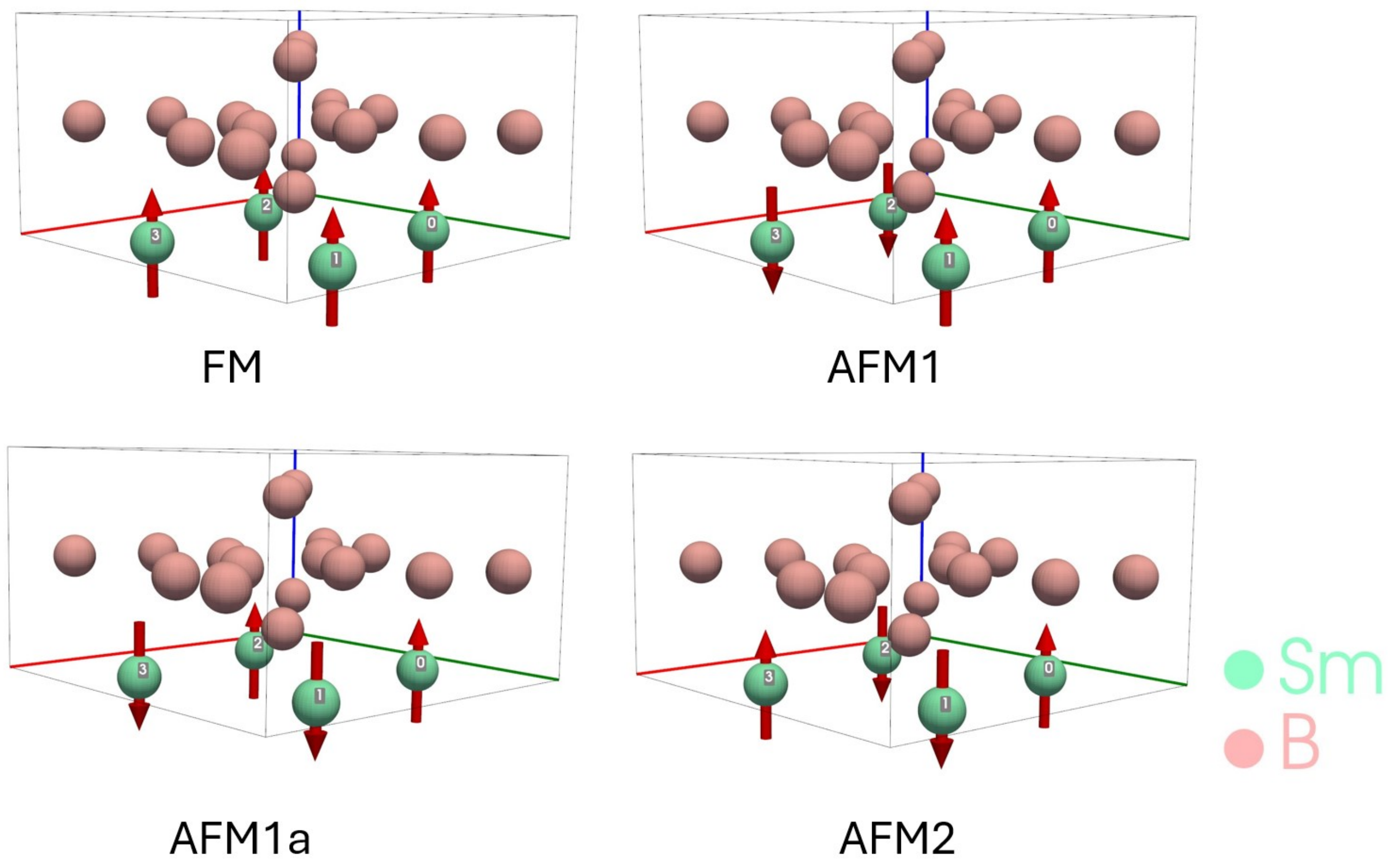

Figure S5: Magnetic configurations of material No. 4, $SmB_4$ (mp-8546), with space group P4/mbm (127). Wyckoff positions: Sm (4g), B (4h, 8j, 4e). Inversion-related Sm pairs: (Sm0, Sm3) and (Sm1, Sm2). No Sm atoms are related by centering translations.

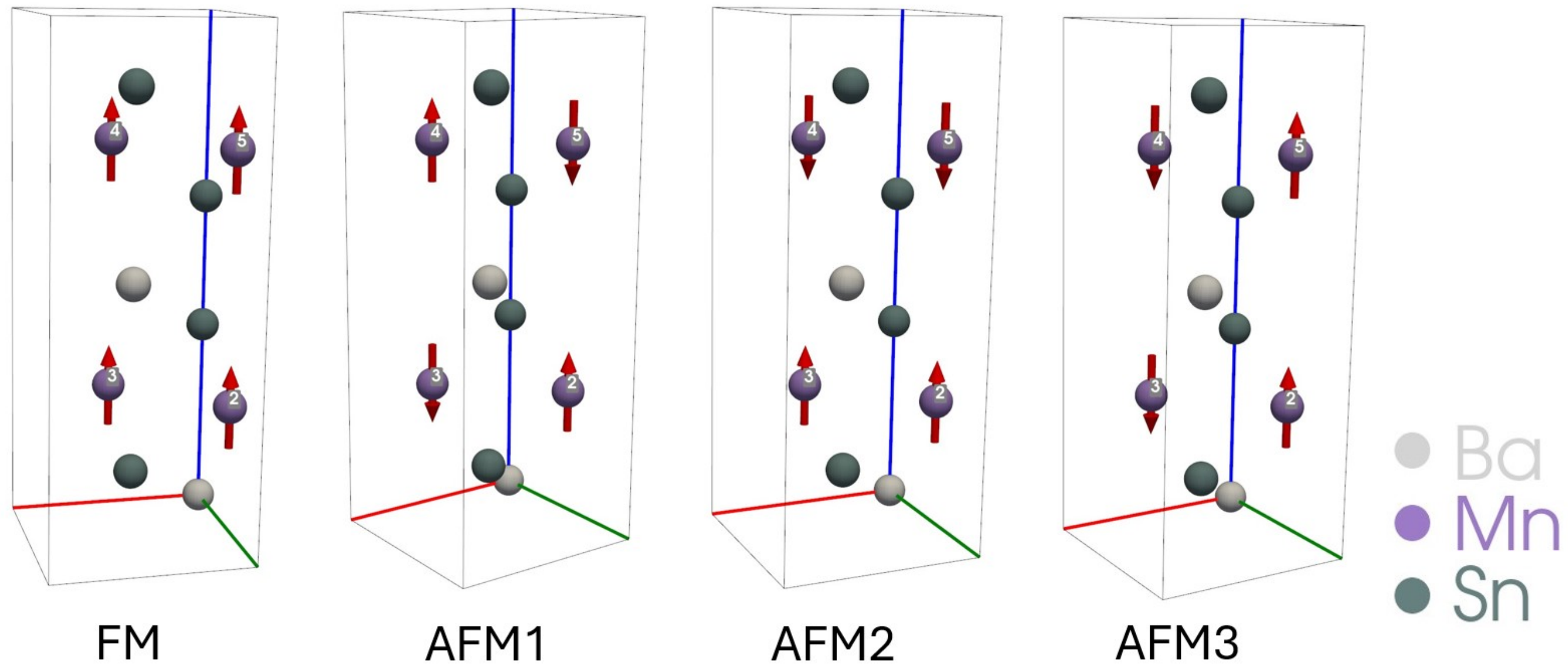

Figure S6: Magnetic configurations of material No. 5, $BaMn_2Sn_2$ (mp-22679), with space group I4/mmm (139). Wyckoff positions: Ba (2a), Mn (4d), Sn (4e). Inversion-related Mn pairs: (Mn2, Mn5) and (Mn3, Mn4). Centering-translation-related Mn pairs: (Mn2, Mn4) and (Mn3, Mn5), with the centering vector (0.5, 0.5, 0.5).

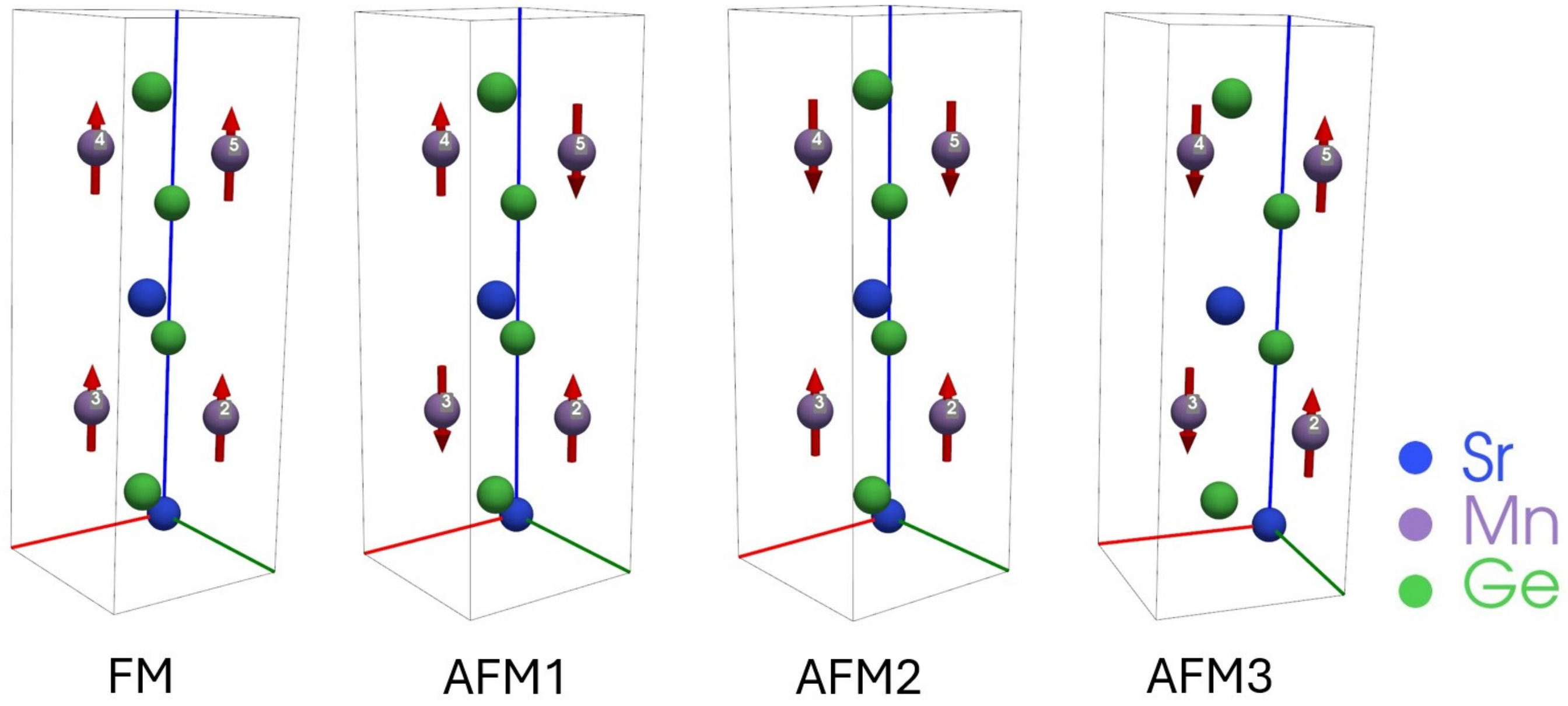

Figure S7: Magnetic configurations of material No. 6, $SrMn_2Ge_2$ (mp-21118), with space group I4/mmm (139). Wyckoff positions: Sr (2a), Mn (4d), Ge (4e). Inversion-related Mn pairs: (Mn2, Mn5) and (Mn3, Mn4). Centering-translation-related Mn pairs: (Mn2, Mn4) and Mn3, Mn5), with the centering vector (0.5, 0.5, 0.5).

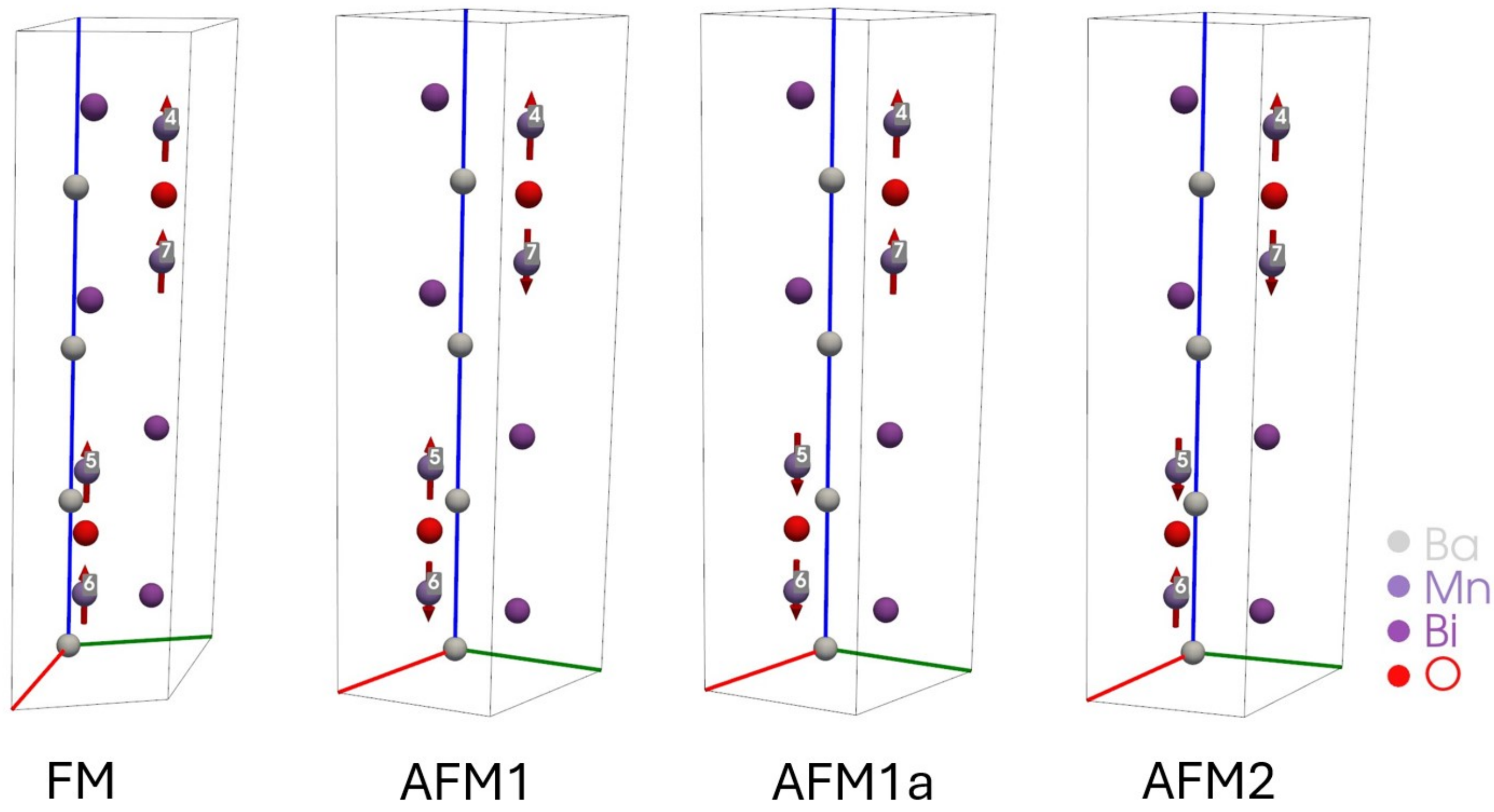

Figure S8: Magnetic configurations of material No. 7, $Ba_2Mn_2Bi_2O$ (mp-556391), with space group $P6_3/mmc$ (194). Wyckoff positions: Ba (2b, 2a), Mn (4f), Bi (4f), O (2d). Inversion-related Mn pairs: (Mn4, Mn6) and (Mn5, Mn7). No Mn atoms are related by centering translations.

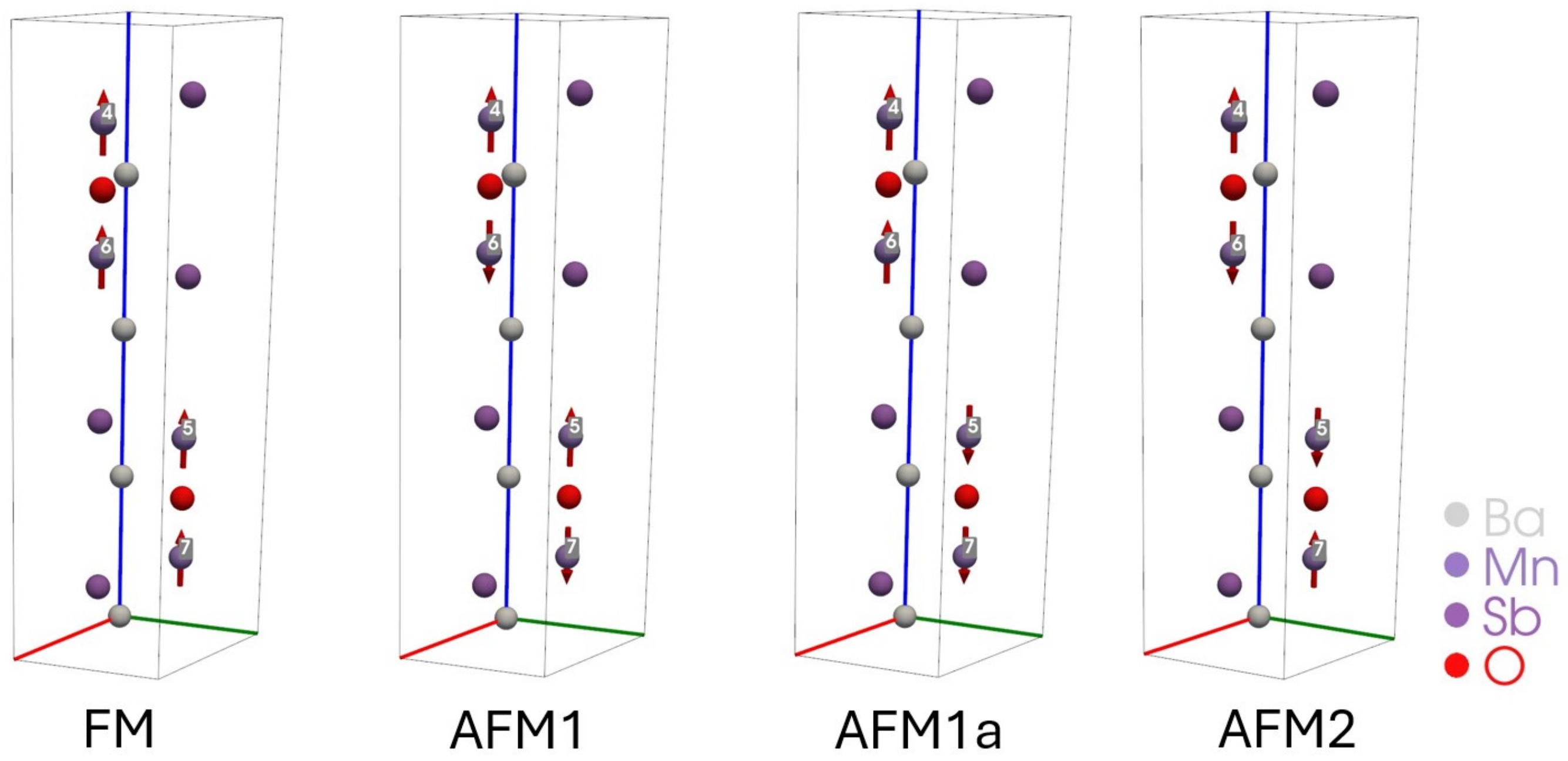

Figure S9: Magnetic configurations of material No. 8, $Ba_2Mn_2Sb_2O$ (mp-19213), with space group $P6_3/mmc$ (194). Wyckoff positions: Ba (2a, 2b), Mn (4f), Sb (4f), O (2c). Inversion-related Mn pairs: (Mn4, Mn7) and (Mn5, Mn6). No Mn atoms are related by centering translations.

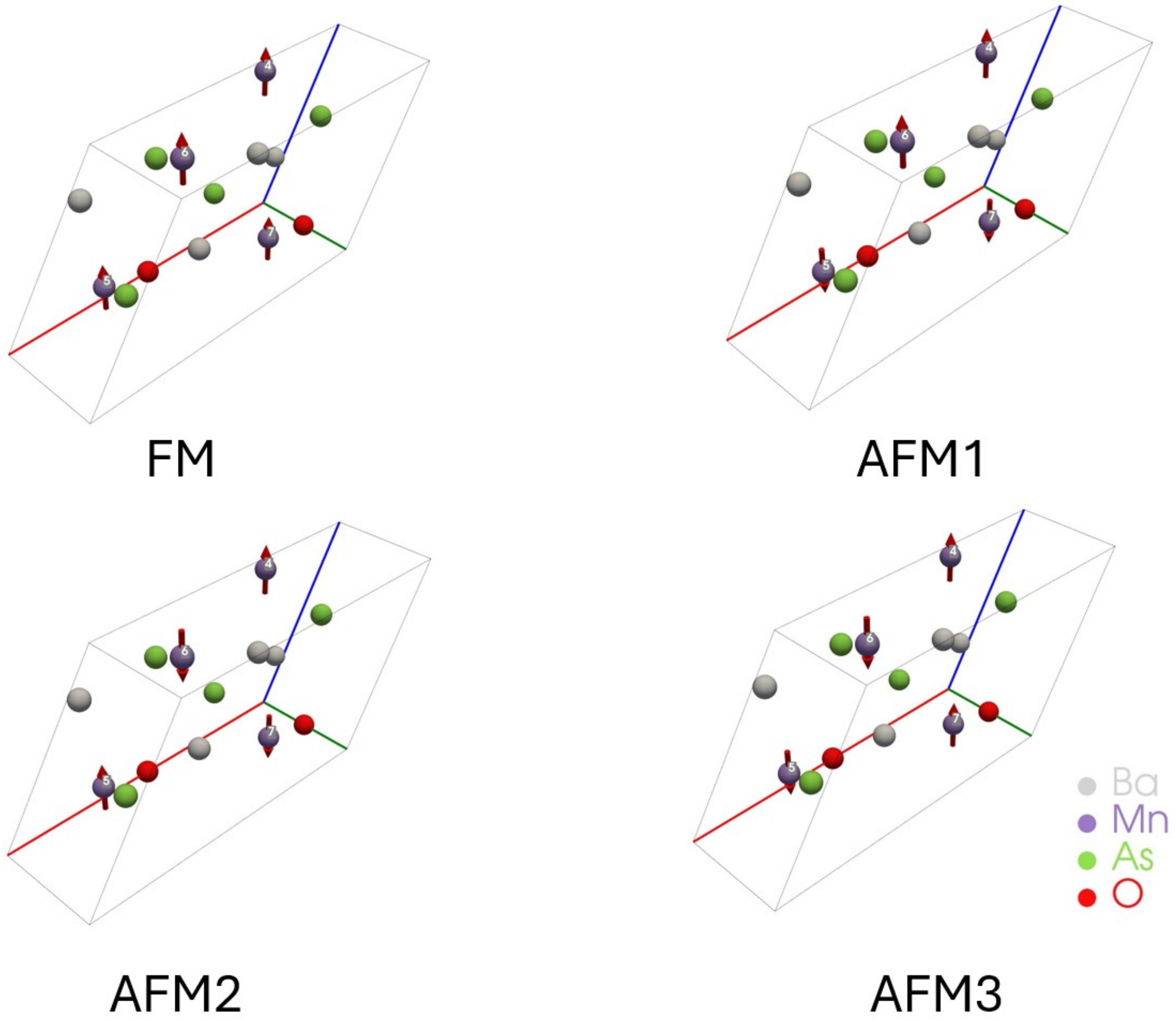

Figure S10: Magnetic configurations of material No. 9, $Ba_2Mn_2As_2O$ (mp-550454), with space group C2/m (12). Wyckoff positions: Ba (4i), Mn (4i), As (4i), O (2b). Inversion-related Mn pairs: (Mn4, Mn5) and (Mn6, Mn7). Centering-translation-related Mn pairs: (Mn4, Mn6) and (Mn5, Mn7), with the centering vector (0.5, 0.5, 0).

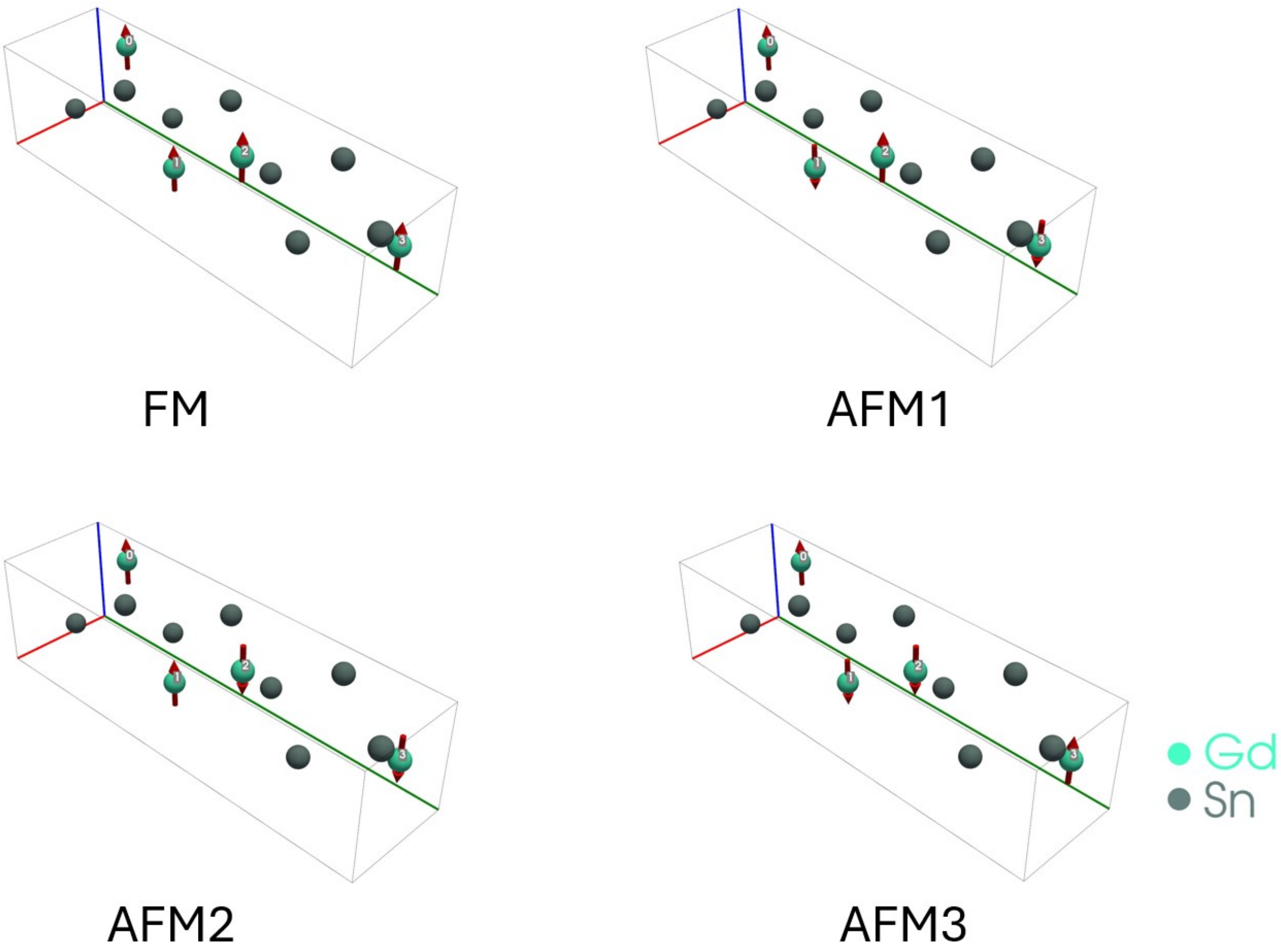

Figure S11: Magnetic configurations of material No. 10, $GdSn_2$ (mp-1071567), with space group Cmcm (63). Wyckoff positions: Gd (4c), Sn (4c, 4c). Inversion-related Gd pairs: (Gd0, Gd3) and (Gd1, Gd2). Centering-translation-related Gd pairs: (Gd0, Gd2) and (Gd1, Gd3), with the centering vector (0.5, 0.5, 0).

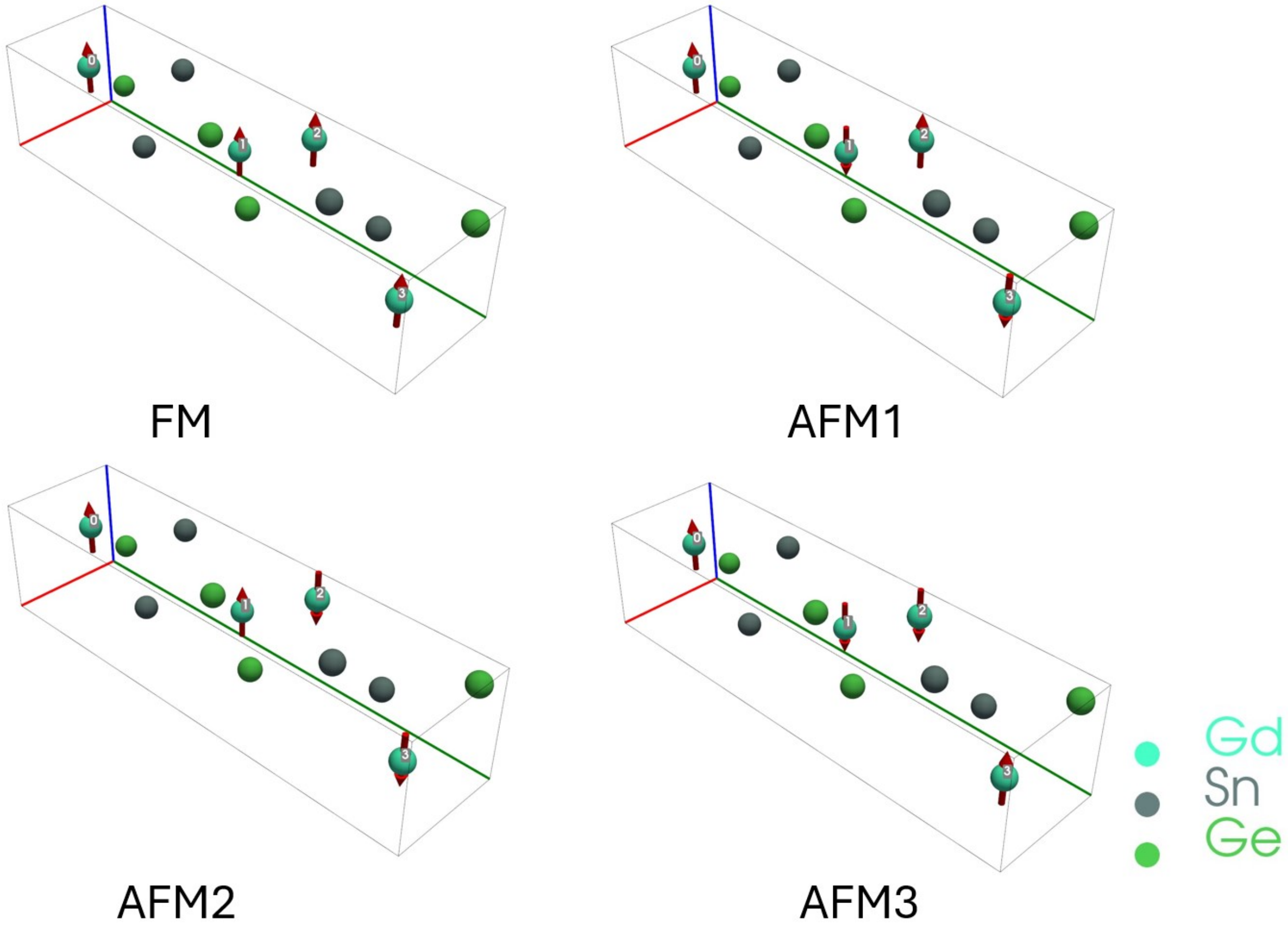

Figure S12: Magnetic configurations of material No. 11, GdSnGe (mp-1206580), with space group Cmcm (63). Wyckoff positions: Gd (4c), Sn (4c), Ge (4c). Inversion-related Gd pairs: (Gd0, Gd3) and (Gd1, Gd2). Centering-translation-related Gd pairs: (Gd0, Gd2) and (Gd1, Gd3), with the centering vector (0.5, 0.5, 0).

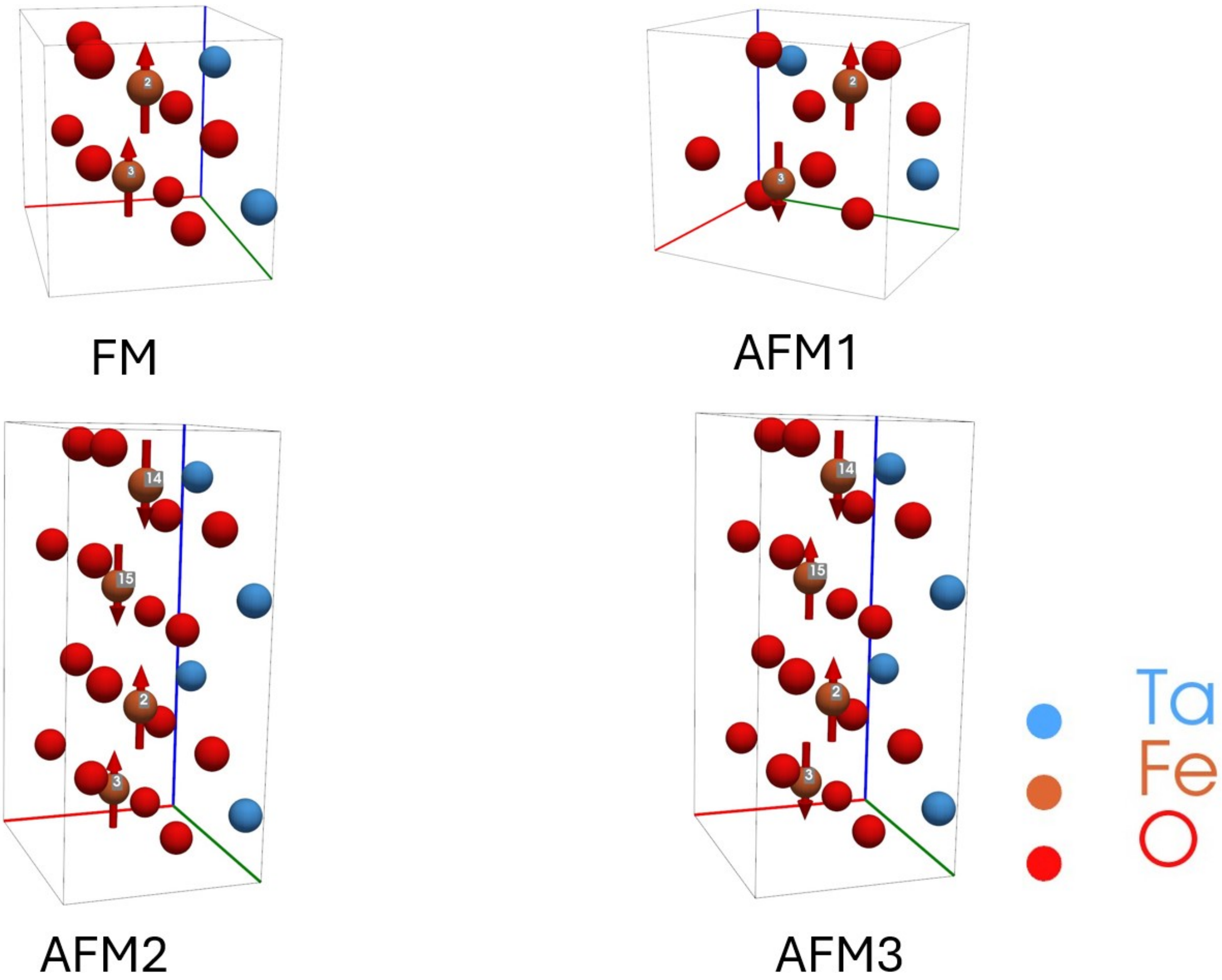

Figure S13: Magnetic configurations of material No. 12, $TaFeO_4$ (mp-755628), with space group P2/c (13). Wyckoff positions: Ta (2e), Fe (2f), O (4g, 4g). Inversion-related Fe pairs: (Fe2, Fe3). No Fe atoms are related by centering translations.

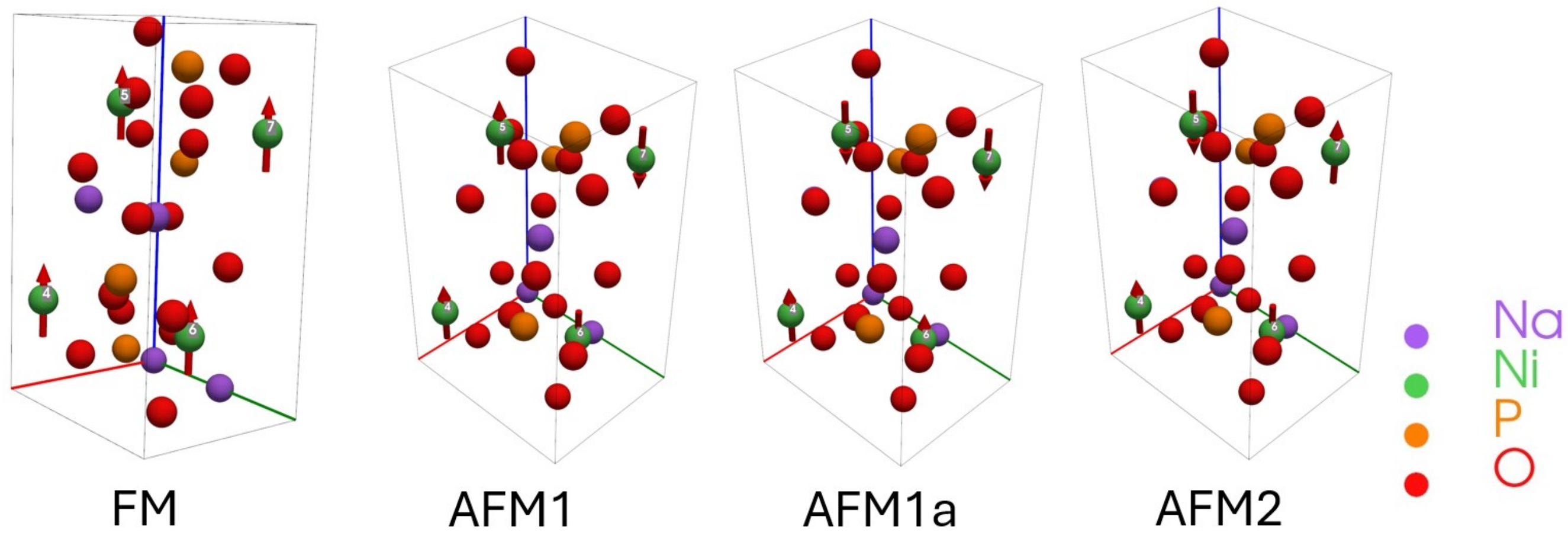

Figure S14: Magnetic configurations of material No. 13, $NaNiPO_4$ (mp-776294), with space group Pnma (62). Wyckoff positions: Na (4a), Ni (4c), P (4c), O (4c, 4c, 8d). Inversion-related Ni pairs: (Ni4, Ni7) and (Ni5, Ni6). No Ni atoms are related by centering translations.

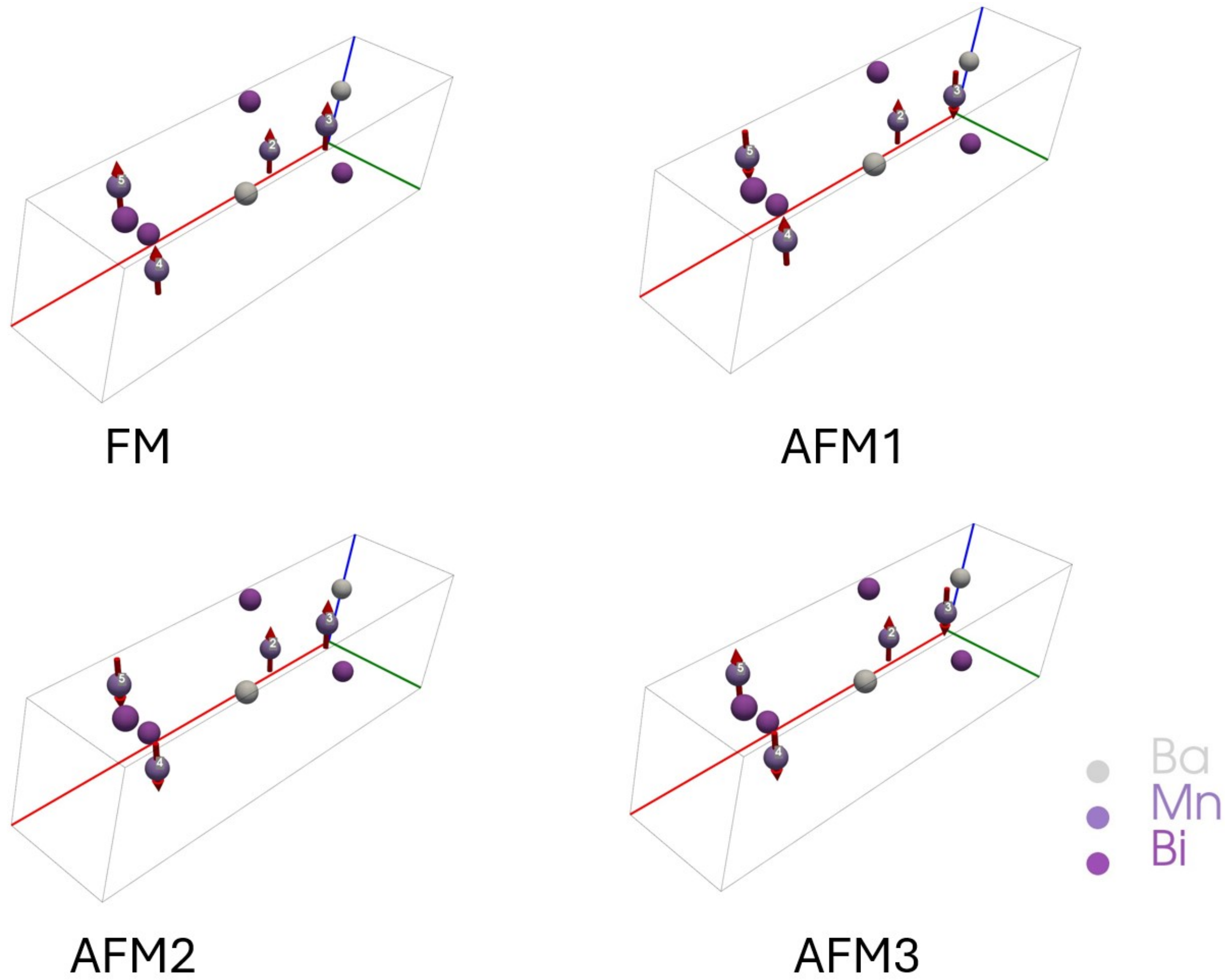

Figure S15: Magnetic configurations of material No. 14, $BaMn_2Bi_2$ (mp-1232615), with space group C2/m (12). Wyckoff positions: Ba (2c), Mn (4i), Bi (4i). Inversion-related Mn pairs: (Mn2, Mn5) and (Mn3, Mn4). Centering-translation-related Mn pairs: (Mn2, Mn4) and (Mn3, Mn5), with the centering vector (0.5, 0.5, 0).

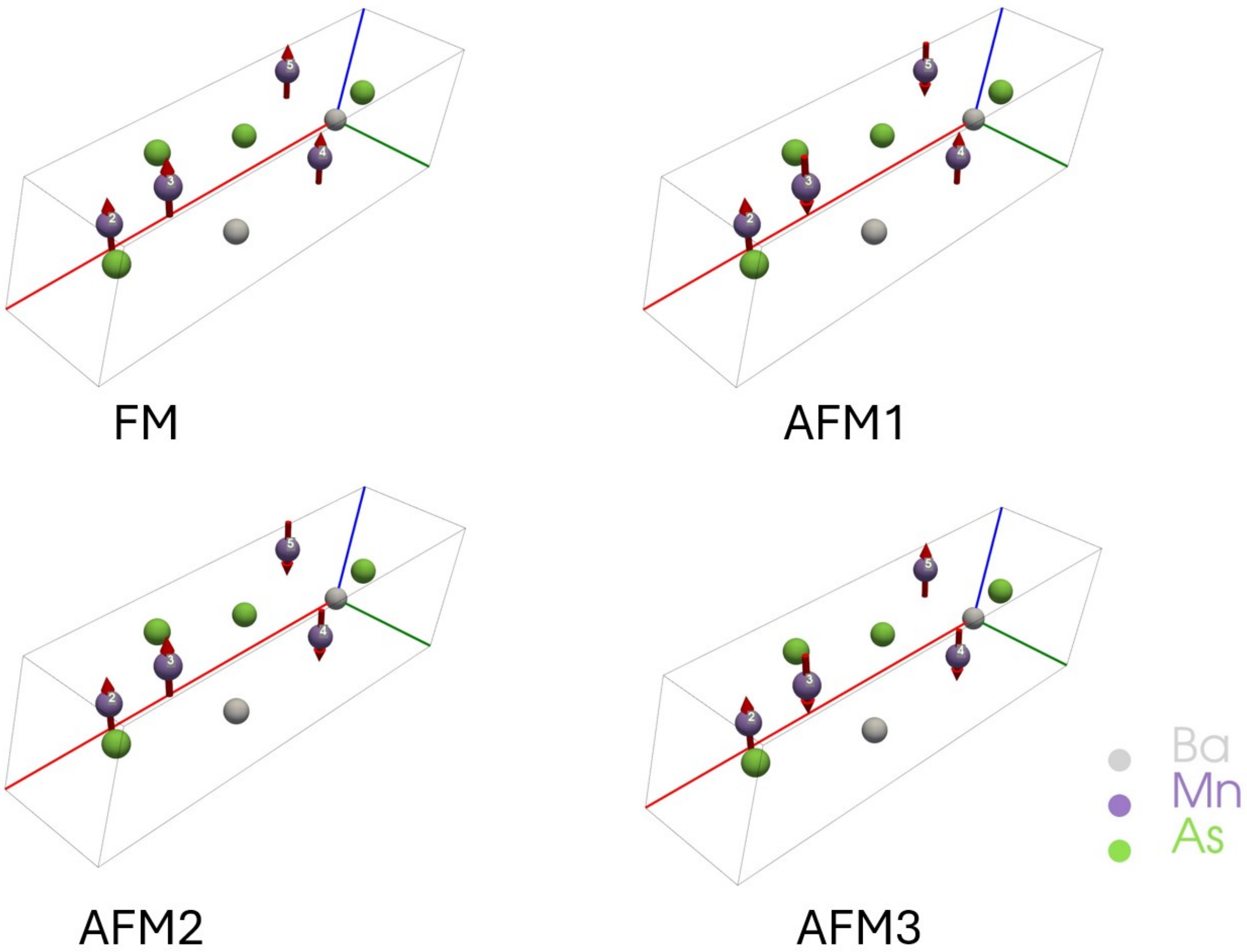

Figure S16: Magnetic configurations of material No. 15, $BaMn_2As_2$ (mp-1232849), with space group C2/m (12). Wyckoff positions: Ba (2a), Mn (4i), As (4i). Inversion-related Mn pairs: (Mn2, Mn5) and (Mn3, Mn4). Centering-translation-related Mn pairs: (Mn2, Mn4) and (Mn3, Mn5), with the centering vector (0.5, 0.5, 0).

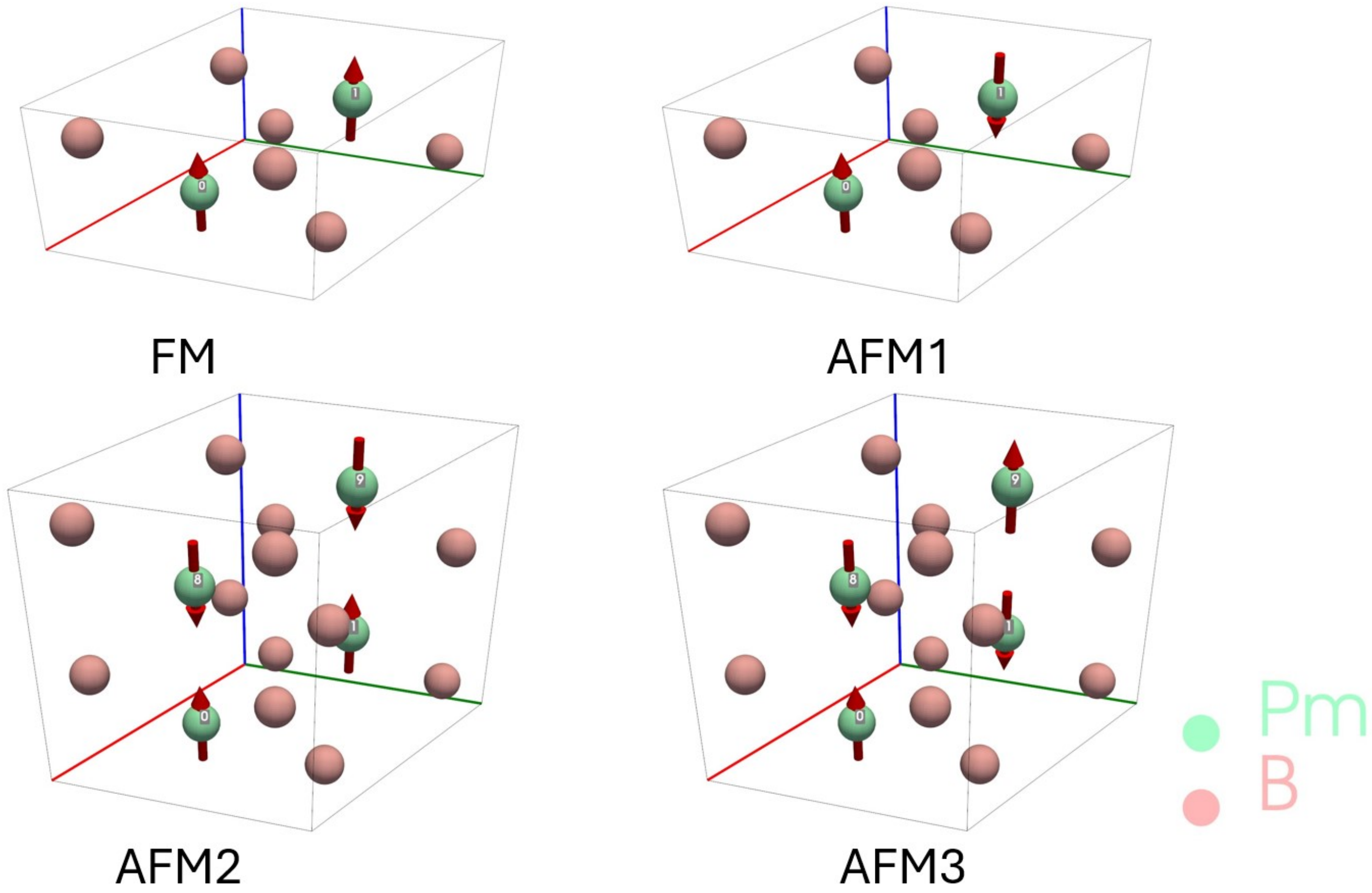

Figure S17: Magnetic configurations of material No. 16, $PmB_3$ (mp-862984), with space group $P6_3/mmc$ (194). Wyckoff positions: Pm (2d), B (6h). Inversion-related Pm pairs: (Pm0, Pm1). No Pm atoms are related by centering translations.

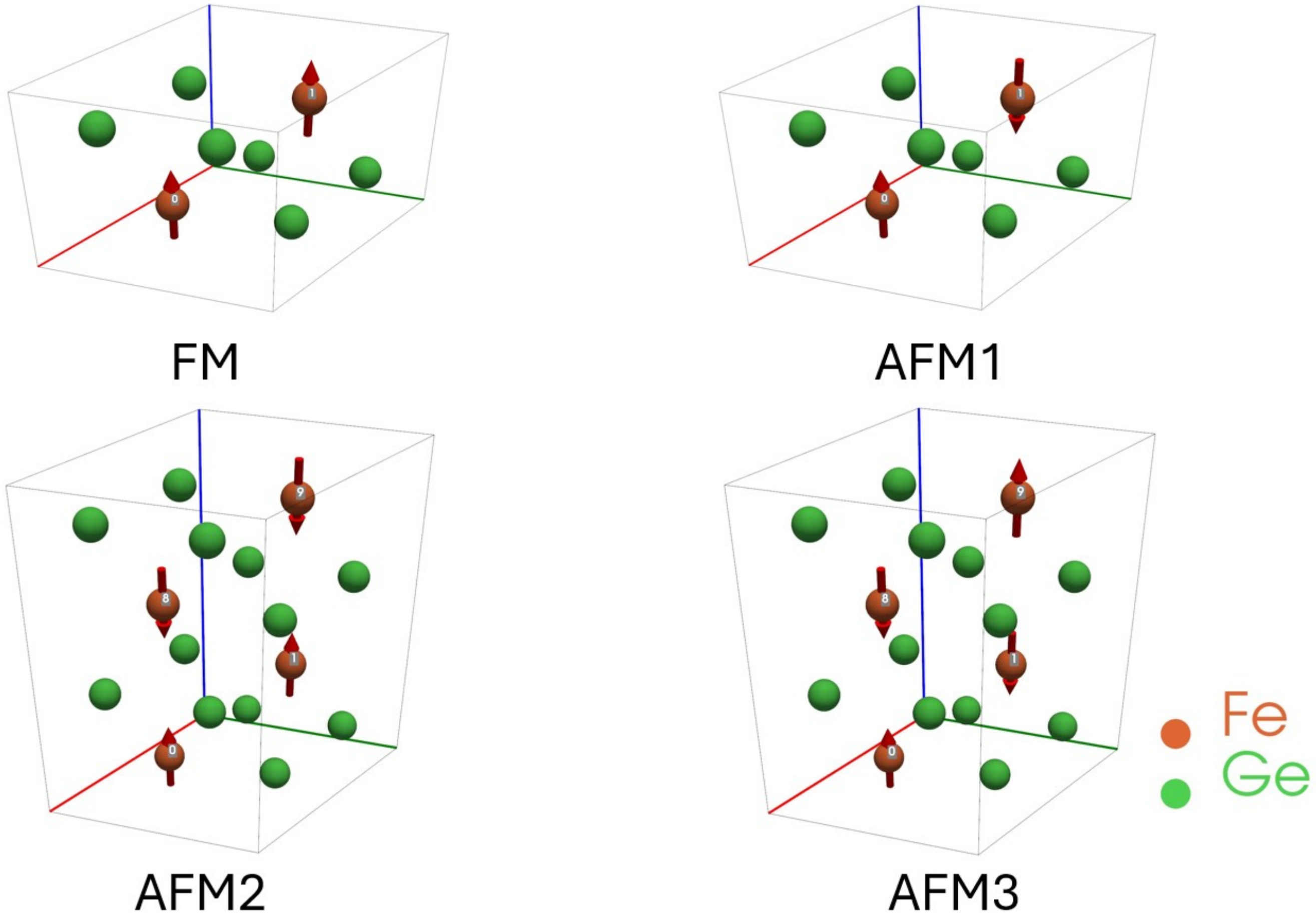

Figure S18: Magnetic configurations of material No. 17, $FeGe_3$ (mp-1184472), with space group $P6_3/mmc$ (194). Wyckoff positions: Fe (2d), Ge (6h). Inversion-related Fe pairs: (Fe0, Fe1). No Fe atoms are related by centering translations.

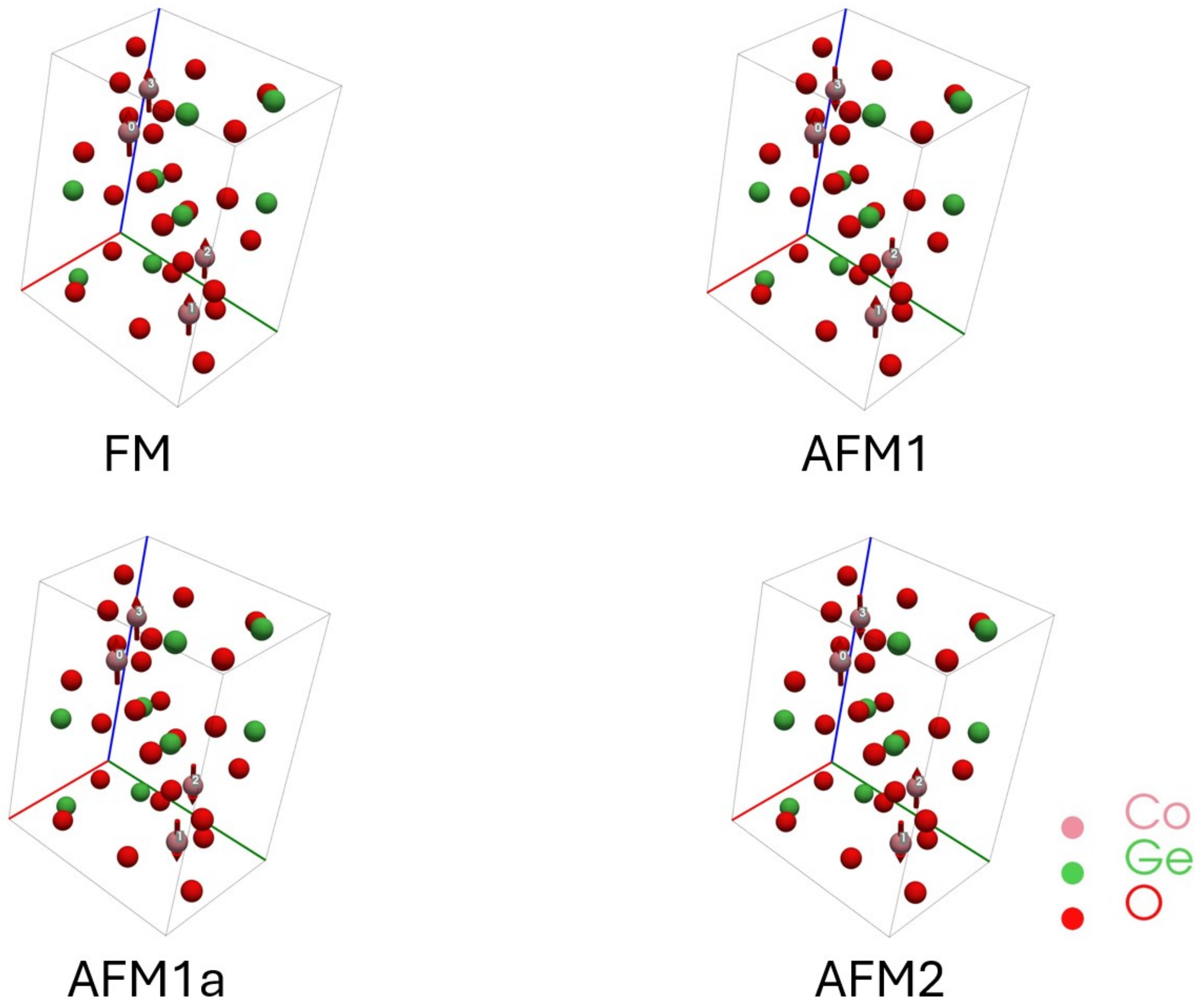

Figure S19: Magnetic configurations of material No. 18, $CoGe_2O_6$ (mp-1353855), with space group $P2_1/c$ (14). Wyckoff positions: Co (4e), Ge (4e, 4e), O (4e, 4e, 4e, 4e, 4e, 4e). Inversion-related Co pairs: (Co0, Co2) and (Co1, Co3). No Co atoms are related by centering translations.

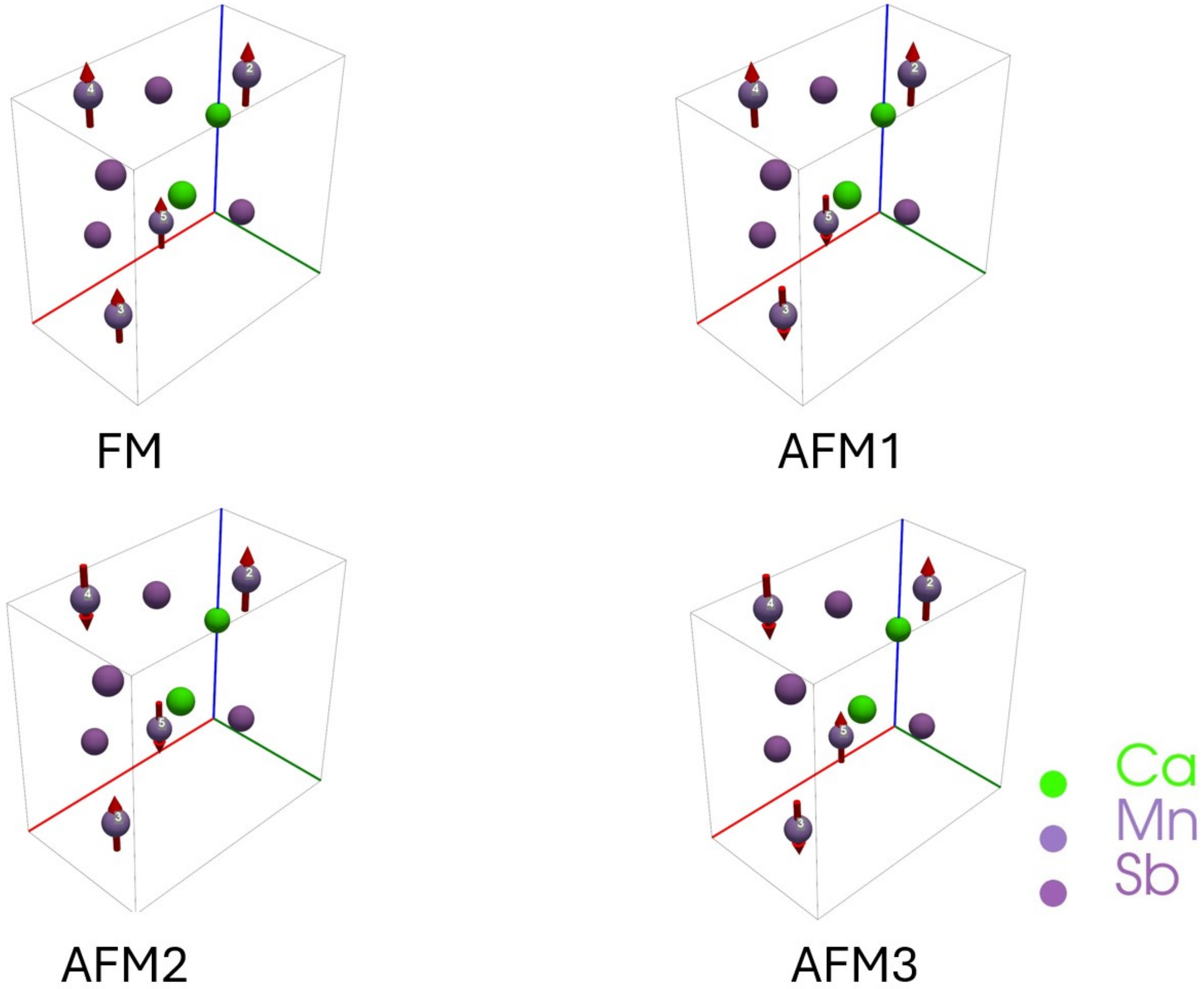

Figure S20: Magnetic configurations of material No. 19, $CaMn_2Sb_2$ (mp-1232722), with space group C2/m (12). Wyckoff positions: Ca (2c), Mn (4i), Sb (4i). Inversion-related Mn pairs: (Mn2, Mn3) and (Mn4, Mn5). Centering-translation-related Mn pairs: (Mn2, Mn4) and (Mn3, Mn5), with the centering vector (0.5, 0.5, 0).

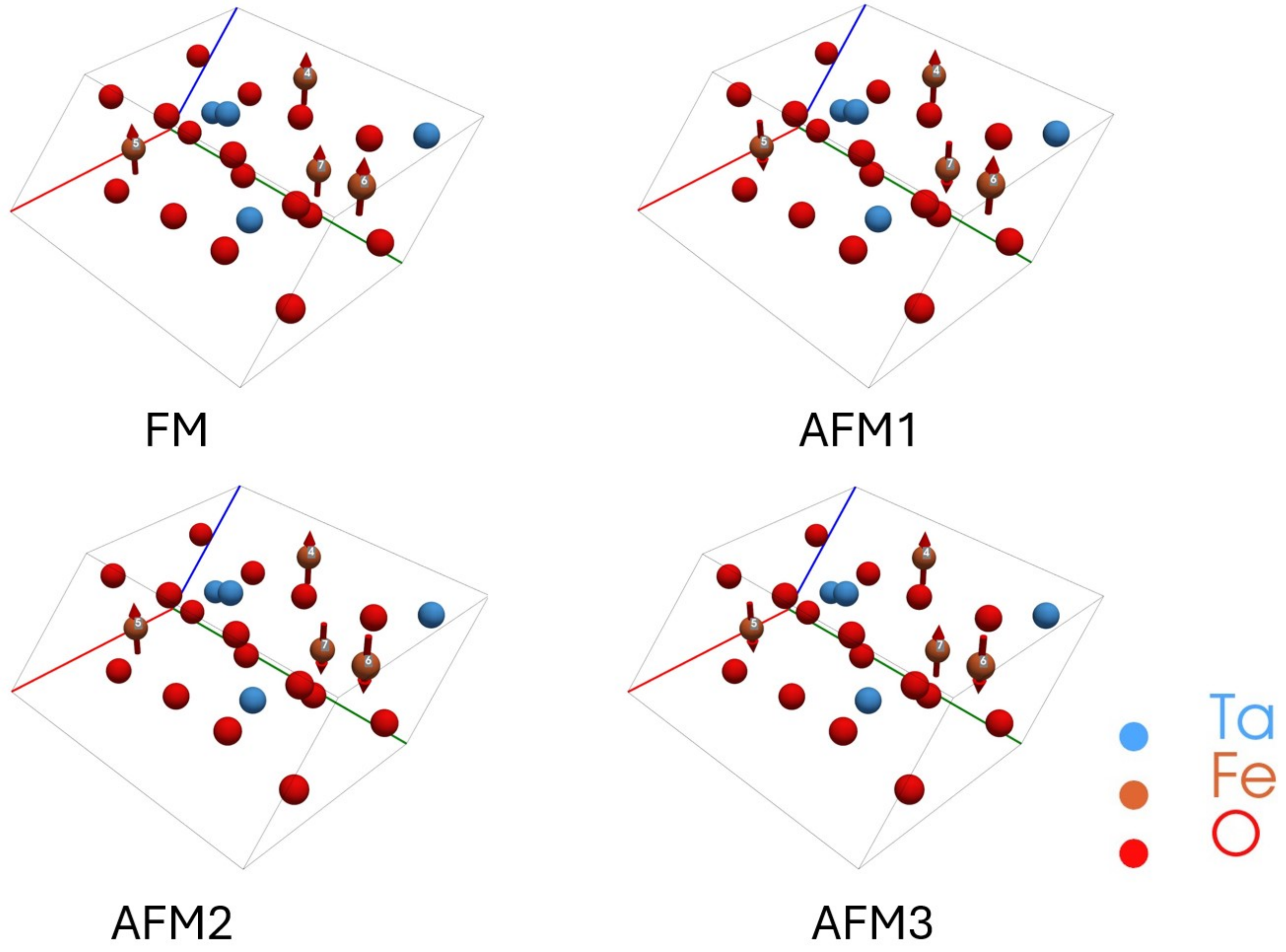

Figure S21: Magnetic configurations of material No. 20, $TaFeO_4$ (mp-755303), with space group C2/c (15). Wyckoff positions: Ta (4e), Fe (4e), O (8f, 8f). Inversion-related Fe pairs: (Fe5, Fe6). Centering-translation-related Fe pairs: (Fe4, Fe6) and (Fe5, Fe7), with the centering vector (0.5, 0.5, 0).

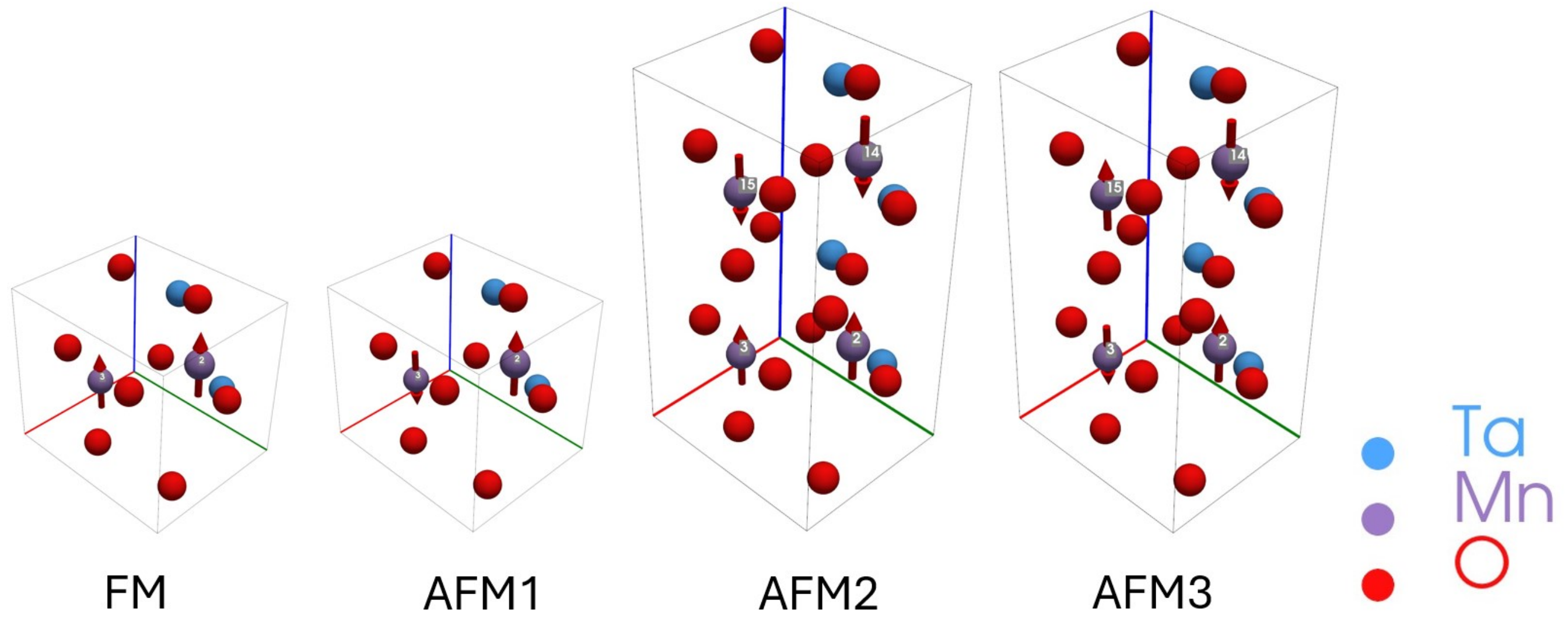

Figure S22: Magnetic configurations of material No. 21, $TaMnO_4$ (mp-1208589), with space group P2/c (13). Wyckoff positions: Ta (2e), Mn (2f), O (4g, 4g). Inversion-related Mn pairs: (Mn2, Mn3). No Mn atoms are related by centering translations.

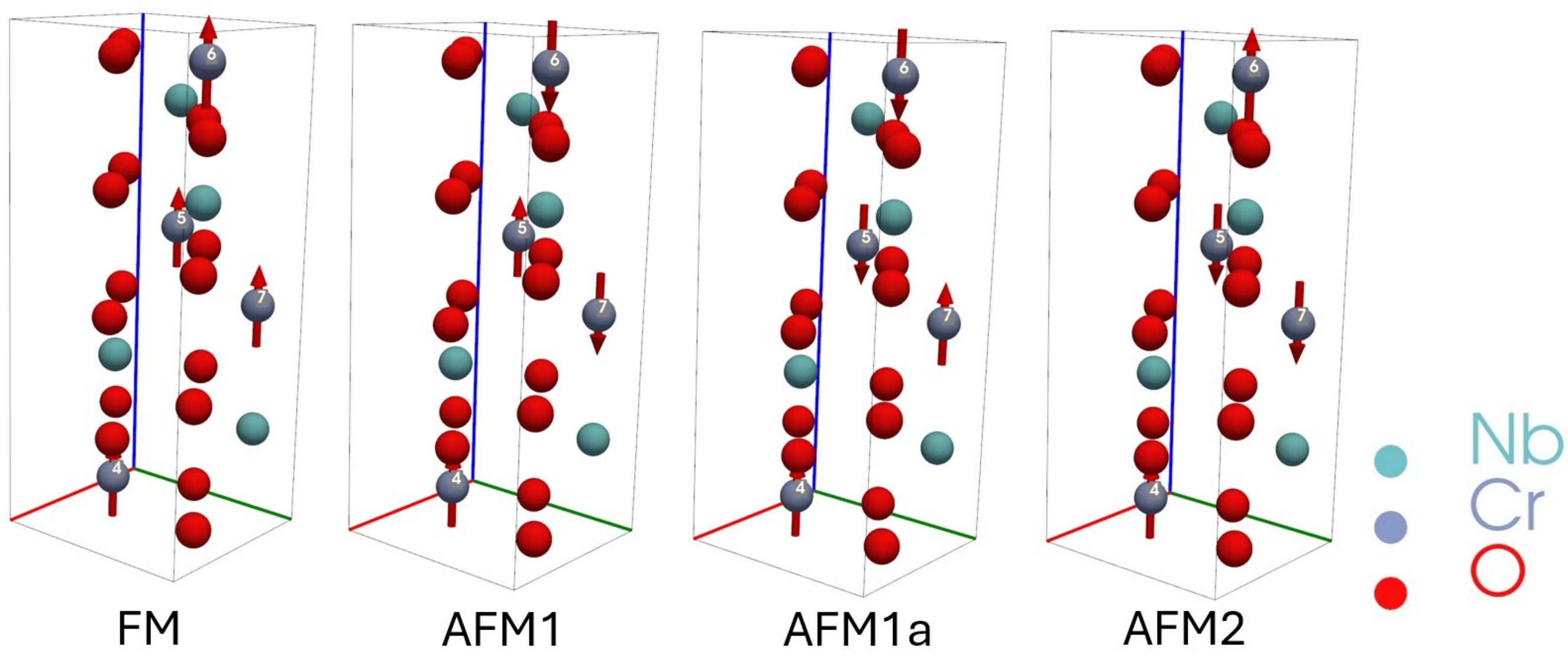

Figure S23: Magnetic configurations of material No. 22, $NbCrO_4$ (mp-772660), with space group Pbcn (60). Wyckoff positions: Nb (4c), Cr (4c), O (8d, 8d). Inversion-related Cr pairs: (Cr4, Cr6) and (Cr5, Cr7). No Cr atoms are related by centering translations.

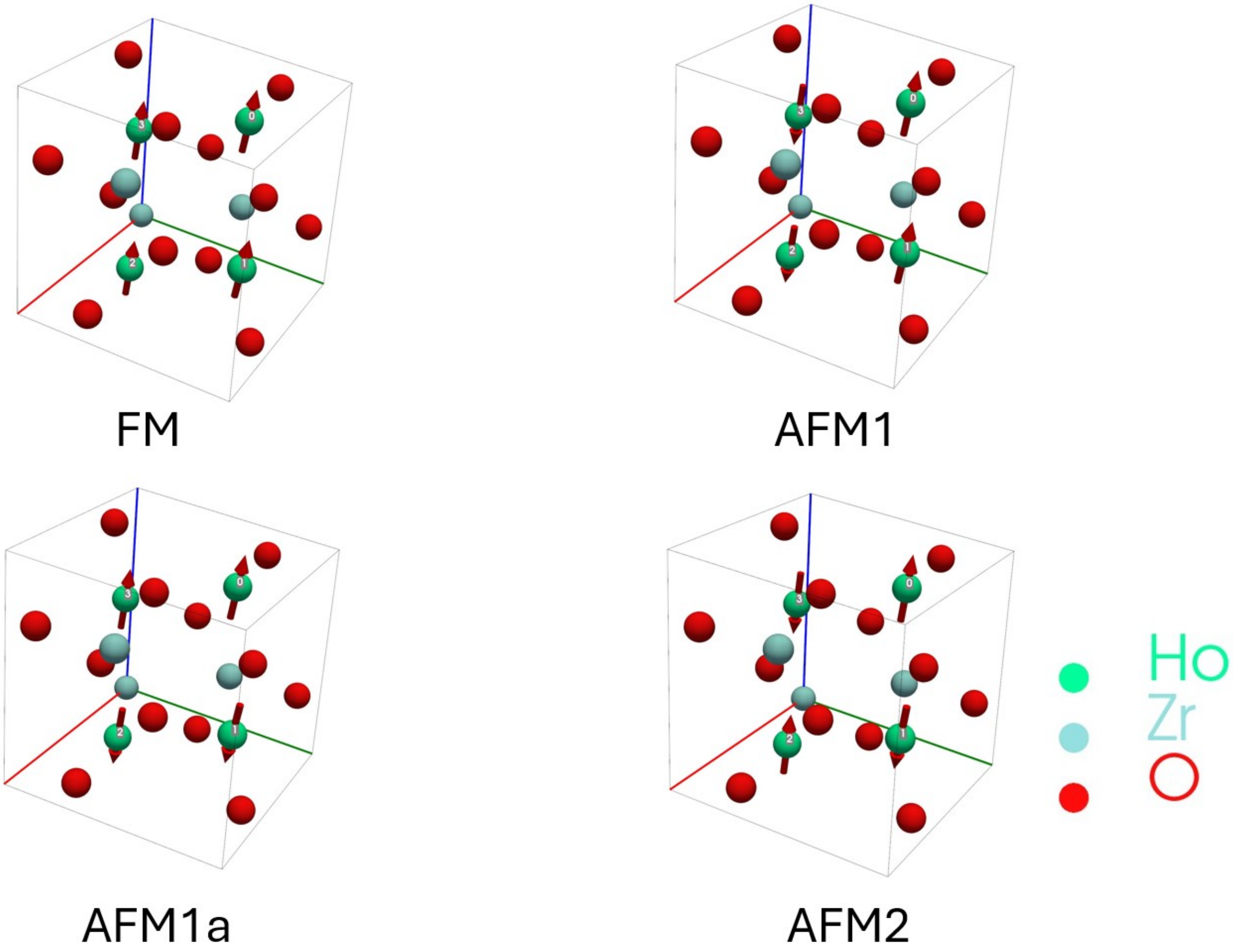

Figure S24: Magnetic configurations of material No. 23, $Ho_4Zr_3O_{12}$ (mp-1224122), with space group $P\overline{1}$ (2). Wyckoff positions: Ho (2i, 2i), Zr (1a), O (2i, 2i, 2i, 2i, 2i, 2i). Inversion-related Ho pairs: (Ho0, Ho2) and (Ho1, Ho3). No Ho atoms are related by centering translations.

# Energy Comparisons of Magnetic Configurations

Table S1: Total energy differences (meV/atom) of different magnetic configurations for the 23 AFM1 candidate materials. The energies are reported relative to the AFM1 configuration, which is set as the reference energy (0). The lowest-energy configuration for each material is highlighted in bold. "N/A" indicates cases where the corresponding magnetic configuration is not applicable. For material No. 2 (mp-513), the FM configuration was computationally found to be 0.09 meV/atom lower in energy than AFM1; however, the material was experimentally confirmed to exhibit AFM1 ordering.

| No. | Material | MP ID | AFM1 | AFM1a | AFM2 | AFM3 | FM |
|---|---|---|---|---|---|---|---|
| 1 | $BaMn_2Ge_2$ | mp-22412 | **0** | N/A | 29.74 | 7.94 | 34.78 |
| 2 | $ErGe_3$ | mp-513 | 0 | N/A | 10.66 | 14.26 | **-0.09** |
| 3 | $NdB_4$ | mp-1632 | 0 | **-2.00** | 7.07 | N/A | 0.01 |
| 4 | $SmB_4$ | mp-8546 | **0** | **0** | 1.43 | N/A | 3.12 |
| 5 | $BaMn_2Sn_2$ | mp-22679 | **0** | N/A | 49.79 | 25.63 | 43.37 |
| 6 | $SrMn_2Ge_2$ | mp-21118 | **0** | N/A | 1.66 | 3.26 | 7.13 |
| 7 | $Ba_2Mn_2Bi_2O$ | mp-556391 | **0** | 22.03 | 3.59 | N/A | 26.25 |
| 8 | $Ba_2Mn_2Sb_2O$ | mp-19213 | **0** | 21.13 | 3.83 | N/A | 25.26 |
| 9 | $Ba_2Mn_2As_2O$ | mp-550454 | **0** | N/A | 23.66 | 69.68 | 97.23 |
| 10 | $GdSn_2$ | mp-1071567 | **0** | N/A | 1.12 | 1.07 | 1.81 |
| 11 | GdSnGe | mp-1206580 | **0** | N/A | 1.73 | 1.39 | 2.44 |
| 12 | $TaFeO_4$ | mp-755628 | **0** | N/A | 2.57 | 2.57 | 8.71 |
| 13 | $NaNiPO_4$ | mp-776294 | **0** | 3.85 | 1.04 | N/A | 4.31 |
| 14 | $BaMn_2Bi_2$ | mp-1232615 | **0** | N/A | 114.02 | 6.56 | 110.43 |
| 15 | $BaMn_2As_2$ | mp-1232849 | **0** | N/A | 157.67 | 9.86 | 156.58 |
| 16 | $PmB_3$ | mp-862984 | **0** | N/A | 16.08 | 2.03 | 10.67 |
| 17 | $FeGe_3$ | mp-1184472 | **0** | N/A | 18.36 | 18.36 | 19.13 |
| 18 | $CoGe_2O_6$ | mp-1353855 | **0** | 3.21 | 1.10 | N/A | 4.52 |
| 19 | $CaMn_2Sb_2$ | mp-1232722 | **0** | N/A | 7.52 | 54.76 | 129.94 |
| 20 | $TaFeO_4$ | mp-755303 | **0** | N/A | 5.68 | 3.42 | 9.93 |
| 21 | $TaMnO_4$ | mp-1208589 | **0** | N/A | 1.90 | 1.90 | 4.43 |
| 22 | $NbCrO_4$ | mp-772660 | **0** | 0.44 | 6.84 | N/A | 7.66 |
| 23 | $Ho_4Zr_3O_{12}$ | mp-1224122 | **0** | 4.96 | 9.38 | N/A | 14.60 |

# Converged Parameters for DFT Calculations

Table S2: Final converged parameters used in DFT calculations. The plane-wave energy cutoff (ecutwfc) and k-point grids were determined through systematic convergence tests to ensure numerical accuracy in total energy comparisons.

| No. | Material | MP ID | ecutwfc (Ry) | **k**-point grid |
|---|---|---|---|---|
| 1 | $BaMn_2Ge_2$ | mp-22412 | 180 | $12 \times 12 \times 12$ |
| 2 | $ErGe_3$ | mp-513 | 180 | $11 \times 11 \times 11$ |
| 3 | $NdB_4$ | mp-1632 | 140 | $12 \times 12 \times 12$ |
| 4 | $SmB_4$ | mp-8546 | 160 | $12 \times 12 \times 12$ |
| 5 | $BaMn_2Sn_2$ | mp-22679 | 180 | $12 \times 12 \times 12$ |
| 6 | $SrMn_2Ge_2$ | mp-21118 | 220 | $12 \times 12 \times 12$ |
| 7 | $Ba_2Mn_2Bi_2O$ | mp-556391 | 200 | $13 \times 13 \times 13$ |
| 8 | $Ba_2Mn_2Sb_2O$ | mp-19213 | 200 | $13 \times 13 \times 13$ |
| 9 | $Ba_2Mn_2As_2O$ | mp-550454 | 180 | $12 \times 12 \times 12$ |
| 10 | $GdSn_2$ | mp-1071567 | 200 | $12 \times 12 \times 12$ |
| 11 | GdSnGe | mp-1206580 | 200 | $11 \times 11 \times 11$ |
| 12 | $TaFeO_4$ | mp-755628 | 200 | $12 \times 12 \times 12$ |
| 13 | $NaNiPO_4$ | mp-776294 | 200 | $10 \times 10 \times 10$ |
| 14 | $BaMn_2Bi_2$ | mp-1232615 | 180 | $14 \times 14 \times 14$ |
| 15 | $BaMn_2As_2$ | mp-1232849 | 180 | $12 \times 12 \times 12$ |
| 16 | $PmB_3$ | mp-862984 | 90 | $14 \times 14 \times 14$ |
| 17 | $FeGe_3$ | mp-1184472 | 180 | $13 \times 13 \times 13$ |
| 18 | $CoGe_2O_6$ | mp-1353855 | 200 | $7 \times 7 \times 7$ |
| 19 | $CaMn_2Sb_2$ | mp-1232722 | 180 | $12 \times 12 \times 12$ |
| 20 | $TaFeO_4$ | mp-755303 | 200 | $8 \times 8 \times 8$ |
| 21 | $TaMnO_4$ | mp-1208589 | 200 | $12 \times 12 \times 12$ |
| 22 | $NbCrO_4$ | mp-772660 | 180 | $13 \times 13 \times 13$ |
| 23 | $Ho_4Zr_3O_{12}$ | mp-1224122 | 120 | $13 \times 13 \times 13$ |